\documentclass[fleqn,usenatbib]{mnras}

\usepackage{newtxtext,newtxmath}

\usepackage[T1]{fontenc}
\usepackage{ae,aecompl}

\usepackage{graphicx}	

\newcommand{\hmsun}{h^{-1}{\rm M}_\odot}
\newcommand{\hmpc}{h^{-1}{\rm Mpc}}

\newcommand{\hubunit}{\rm {km}~{Mpc^{-1}}~{s^{-1}} }
\newcommand{\kms}{\rm {km/s} }

\newcommand{\velv}{\textbf{V}_{\rm v}}
\newcommand{\posv}{\textbf{X}_{\rm v}}
\newcommand{\possvx}{X_{\rm v1}}
\newcommand{\possvy}{X_{\rm v2}}
\newcommand{\possvz}{X_{\rm v3}}

\newcommand{\velvx}{V_{\rm v1}}
\newcommand{\velvy}{V_{\rm v2}}
\newcommand{\velvz}{V_{\rm v3}}

\newcommand{\zsim}{z}

\newcommand{\rrs}{R_{\rm v}^{\rm rs}}
\newcommand{\rzs}{R_{\rm v}^{\rm zs}}

\newcommand{\qrsd}{q_{\rm RSD}}

\newcommand{\qap}{q_{\rm AP}}

\newcommand{\xipos}{\xi_{\rm pos}(\sigma)}
\newcommand{\xiposs}{\xi_{\rm pos}(r_\perp)}
\newcommand{\xilos}{\xi_{\rm los}(\pi)}
\newcommand{\xiloss}{\xi_{\rm los}(r_\parallel)}

\newcommand{\xisp}{\xi(\sigma,\pi)}

\newcommand{\zxz}{z \times z}
\newcommand{\rxz}{r \times z}
\newcommand{\rxr}{r \times r}

\title[Redshift-space effects in voids II]{Redshift-space effects in voids and their impact on cosmological tests.\\
Part II: the void-galaxy cross-correlation function}

\author[Correa et al.]{
Carlos M. Correa,$^{1,2}$\thanks{E-mail: cmcorrea@unc.edu.ar (CMC)}
Dante J. Paz,$^{1,2}$
Nelson D. Padilla$^{3,4}$
Ariel G. S\'anchez,$^{5}$
\newauthor
Andr\'es N. Ruiz,$^{1,2}$
and Ra\'ul E. Angulo$^{6,7}$
\\
$^{1}$Instituto de Astronom\'ia Te\'orica y Experimental, UNC-CONICET, Laprida 854, X5000BGR C\'ordoba, Argentina\\
$^{2}$Observatorio Astron\'omico de C\'ordoba, Universidad Nacional de C\'ordoba, Laprida 854, X5000BGR C\'ordoba, Argentina\\
$^{3}$ Instituto de Astrof\'isica, Pontificia Universidad Cat\'olica de Chile, Av. Vicu\~na Mackenna 4860, Santiago, Chile\\
$^{4}$ Centro de Astro-ingenier\'ia, Pontificia Universidad Cat\'olica de Chile, Av. Vicu\~na Mackenna 4860, Santiago, Chile\\
$^{5}$Max-Planck-Institut f\"ur Extraterrestrische Physik, Postfach 1312, Giessenbachstr, D-85741 Garching, Germany\\
$^{6}$ Donostia International Physics Centre (DIPC), Paseo Manuel de Lardizabal 4, E-20018 Donostia-San Sebastian, Spain\\
$^{7}$ IKERBASQUE, Basque Foundation for Science, E-48013 Bilbao, Spain\\
}

\date{Accepted XXX. Received YYY; in original form ZZZ}

\pubyear{2021}

\begin{document}
\label{firstpage}
\pagerange{\pageref{firstpage}--\pageref{lastpage}}
\maketitle


\begin{abstract}
This is the second part of a thorough investigation of the redshift-space effects that affect void properties and the impact they have on cosmological tests.
Here, we focus on the void-galaxy cross-correlation function, specifically, on the projected versions that we developed in a previous work.
The pillar of the analysis is the one-to-one relationship between real and redshift-space voids above the shot-noise level identified with a spherical void finder.
Under this mapping, void properties are affected by three effects: (i) a systematic expansion as a consequence of the distortions induced by galaxy dynamics, (ii) the Alcock-Paczynski volume effect, which manifests as an overall expansion or contraction depending on the fiducial cosmology, and (iii) a systematic off-centring along the line of sight as a consequence of the distortions induced by void dynamics.
We found that correlations are also affected by an additional source of distortions: the ellipticity of voids.
This is the first time that distortions due to the off-centring and ellipticity effects are detected and quantified.
With a simplified test, we verified that the Gaussian streaming model is still robust provided all these effects are taken into account, laying the foundations for improvements in current models in order to obtain unbiased cosmological constraints from spectroscopic surveys.
Besides this practical importance, this analysis also encodes key information about the structure and dynamics of the Universe at the largest scales.
Furthermore, some of the effects constitute cosmological probes by themselves, as is the case of the void ellipticity.
\end{abstract}




\section{Introduction}
\label{sec:intro}

Cosmic voids are promising cosmological probes for testing the dark energy problem and alternative gravity theories.
Since they are the largest observable structures, they encode key information about the geometry and expansion history of the Universe.
The potential of voids has been increased recently with the development of modern spectroscopic surveys, such as the Baryon Oscillation Spectroscopic Survey \citep[BOSS]{boss}, its extension eBOSS \citep{eboss}, and the future Dark Energy Spectroscopic Instrument \citep[DESI]{desi}, the Euclid mission \citep{euclid}, and the Hobby-Eberly Telescope Dark Energy Experiment \citep[HETDEX]{hetdex}, which will cover a volume and redshift range without precedents.
In view of this, it will be possible to obtain rich samples of voids at different redshifts, and in this way, to test the evolution of the Universe with high precision.

Two of the most important cosmological statistics in void studies are the void size function, that describes the abundance of voids \citep{svdw, abundance_furlanetto, jennings_Vdn, abundance_achitouv, abundance_pisani, abundance_ronconi1, abundance_bias_contarini, abundance_ronconi2, abundance_verza}, and the void-galaxy cross-correlation function, that characterises the density and peculiar velocity fields around them \citep{clues2, aprsd_hamaus1, rsd_cai, aprsd_hamaus2, rsd_achitouv1, rsd_achitouv2, rsd_chuang, rsd_hamaus, rsd_hawken1, rsd_achitouv3, rsd_nadathur, aprsd_nadathur, reconstruction_nadathur, rsd_hawken2, aprsd_hamaus2020, aprsd_nadathur2020, rsd_paillas}.
Both statistics are affected by distortions in the observed spatial distribution of the galaxies, which translate into anisotropic patterns.
There are two main sources of distortions: the redshift-space distortions \citep[hereafter RSD]{kaiser_rsd} and the \citet[AP]{ap} effect.
The former is a dynamical effect, caused by the impact of the peculiar velocities of galaxies on the measured redshifts, whereas the latter, a geometrical effect, caused by the fiducial cosmology used to transform the angles and redshifts from a survey into distances expressed in a physical scale.
These distortions can be modelled from physical principles, and therefore, they encode fundamental information about the cosmological parameters.

Nevertheless, our standard picture of distortions around voids is incomplete.
\cite{rsd_nadathur} and \cite{reconstruction_nadathur} noticed that there are four commonly assumed hypotheses in all RSD models for voids, which are violated when voids are identified in redshift space, i.e. from observations.
These postulates that, under the redshift-space mapping:
\begin{enumerate}
    \item the number of void-galaxy pairs must be conserved;
    \item the position of void centres must remain invariant;
    \item the density field around voids must still be radially directed (from the point of view of the galaxies in real space);
    \item the peculiar velocity field must also be isotropic.
\end{enumerate}
The failure of these hypotheses finds an explanation in the RSD and AP effects, which besides distorting the spatial distribution of the galaxies around voids, they also have a direct impact on the void identification process itself, affecting intrinsic void properties, such as their number, size and spatial distribution.
This is because voids are searched for and found using galaxy redshifts and angular positions.
This problematic is important when designing cosmological tests, since these void systematics generate deviations on the observations that lead to biased cosmological constraints if they are not taken into account properly.
The validity of these hypotheses is also discussed in \cite{aprsd_hamaus2020}. 

One approach to avoid these systematics is to use a reconstruction technique \citep{reconstruction_eisenstein}, which has been first applied to baryonic acoustic oscillations analyses.
The method consists of recovering the real-space position of the galaxies based on the Zel'dovich approximation.
This method has now been applied to the case of voids \citep{aprsd_nadathur, aprsd_nadathur2020}, showing robustness in recovering the real-space statistical properties of voids, such as their density and velocity profiles, and achieving unbiased cosmological constraints.
Given that this procedure depends on the cosmological parameters, it must be applied iteratively in combination with the void finding step.
Therefore, it also has some disadvantages.
For instance, this results in an increment of the computational cost of model evaluation.
Moreover, it is quite redundant, since it removes the peculiar velocity affectation of galaxies in order to find the optimal centres of voids, but to achieve high precision constraints, then it must be combined with an RSD analysis of the galaxies around these realistic voids.
Finally, as we shall demonstrate in this work, there are valuable cosmological and dynamical information contained in the redshift-space void systematics that are not fully exploited if the reconstruction technique is employed.

In \citet[hereafter Paper~I]{zvoids_correa}, we proposed an alternative approach, namely, to find a physical connection between the identification of voids in real space (hereafter $r$-space) and in redshift space ($z$-space) using a spherical void finder.
The case of voids is more complex than the case of galaxies.
Galaxies, on the one hand, can be considered as particles, which are totally conserved under the $z$-space mapping, only their position changes.
Voids, on the other hand, are extensive regions of space, hence some of them can be destroyed under this mapping, whereas other artificial ones can be created.
Moreover, the properties of voids are sensitive to the method employed for their identification.
Therefore, it is not clear a priori if there is a relation between both void populations.
We found three relevant results.
First, voids above the shot-noise level are almost conserved under the $z$-space mapping; the void loss decreases as larger voids are considered.
In view of this, it is valid to assume void number conservation when modelling.
Second, two independent effects act on the volume of voids under this mapping.
One is a systematic expansion, a by-product of the RSD induced by tracer dynamics.
The other is an overall expansion or contraction due to the AP effect, depending on the fiducial cosmology.
Third, the position of the void centres are systematically shifted along the line-of-sight direction (hereafter LOS), a by-product of a different class of RSD induced at larger scales by the global dynamics of the whole regions containing the voids, which can be considered as a void dynamics \citep{lambas_sparkling_2016,ceccarelli_sparkling_2016,lares_sparkling_2017}.
We provided a theoretical framework to describe physically these effects from dynamical and cosmological considerations.

In Paper~I we analysed the impact that these void systematics have on the void size function as a cosmological test.
The present work is a continuation of Paper~I, focusing now on the void-galaxy cross-correlation function.
To carry out this study, we adopted the methodology of \citet[hereafter C19]{aprsd_correa} by analysing two perpendicular projections of the correlation function with respect to the LOS.
Using these projections have some advantages.
For instance, they are affected by RSD and AP distortions in different proportions.
Moreover, they can be measured directly in terms of void-centric angular distances and redshifts without the need to assume a fiducial cosmology.
These features allow us to break effectively the degeneracies between the geometrical and dynamical distortions.
Furthermore, our model takes into account a third source of systematics besides the AP and RSD effects: the mixture of scales due to the binning scheme.
This is what allowed us to work with full projections.
Finally, the associated data covariance matrices are much smaller than those corresponding to the traditional way of measuring correlations, and the noise in the likelihood analysis is reduced, which allows to use a smaller number of mock catalogues to measure them.

This paper is organised as follows.
In Section~\ref{sec:data}, we describe the data set, that is, the numerical N-body simulation and the void catalogues.
We also describe the bijective mapping between voids in real and redshift space, along with the sample selected to measure the correlation function.
In Section~\ref{sec:correlation}, we review and adapt our method based on the projected correlation functions from C19, and present the results from the simulation.
In Section~\ref{sec:zeffects}, we recapitulate the description of the $z$-space effects that affect void properties from Paper~I.
Section~\ref{sec:impact} constitutes the core of this paper, where we explain in full detail the impact of these $z$-space effects on the projected correlations and how to account for them based on the theoretical framework of the previous section.
Finally, we summarise and discuss our results in Section~\ref{sec:conclusions}.

\section{Data set}
\label{sec:data}

\subsection{Simulation setup}
\label{subsec:data_sim}

We used the same data set as in C19 and Paper~I.
Briefly, we used the Millennium XXL N-body simulation \citep[hereafter MXXL]{mxxl}, which follows the evolution of $6720^3$ dark matter particles inside a cubic box of length $3000~\hmpc$.
The particle mass is $8.46\times10^9~\hmsun$ in a flat-$\Lambda$CDM cosmology with the following cosmological parameters: $\Omega_m=0.25$, $\Omega_\Lambda=0.75$, $\Omega_b=0.045$, $\Omega_\nu=0.0$, $h=0.73$\footnote{The Hubble constant is parametrised as $H_0=100~h~\hubunit$. All distances and masses are expressed in units of $\hmpc$ and $\hmsun$, respectively.}, $n_s=1.0$ and $\sigma_8=0.9$.
We used the snapshot belonging to redshift $\zsim=0.99$, assumed as the mean redshift of the sample.

Dark matter haloes were chosen as tracers, which were identified as groups of more than $60$ particles using a friends-of-friends algorithm with a linking length parameter of $0.2$ times the mean inter particle separation.
We selected a lower mass cut of $5\times10^{11}~\hmsun$, finding in this way $136,688,808$ haloes.

Positions $\mathbf{x} = (x_1,x_2,x_3)~[\hmpc]$ and peculiar velocities $\mathbf{v} = (v_1,v_2,v_3)~[\kms]$ of haloes in real space were available to quantify the effects of distortions.
In order to generate RSD, we treated the $x_3$-axis of the simulation box as the LOS direction, assuming the distant observer approximation.
We applied the following equation to shift the LOS coordinates of haloes from real to redshift space:
\begin{equation}
    \Tilde{x}_3 = x_3 + v_3 \frac{(1 + \zsim)}{H(\zsim)},
	\label{eq:halo_zspace}
\end{equation}
where $\Tilde{x}_3$ denotes the shifted $x_3$-coordinate, and $H(z)$ is the Hubble parameter, which for a flat-$\Lambda$CDM cosmology can be expressed in terms of the cosmological parameters as follows:
\begin{equation}
    H(z) = 100~h~\sqrt{\Omega_m(1+z)^3 + \Omega_\Lambda},
    \label{eq:hubble}
\end{equation}
where in turn, $\Omega_\Lambda = 1 - \Omega_m$.

\subsection{Void catalogues}
\label{subsec:data_voids}

A detailed description of the void catalogues used in this work can be found in Paper~I.
Briefly, we applied the spherical void finder developed by \citet{clues3}, which is a modified version of the algorithm of \citet{voids_padilla}.
Below, we summarise the main steps.

\begin{enumerate}

\item
\textit{Voronoi tessellation.}
A Voronoi tessellation is performed to obtain an estimation of the density field.
We used a parallel version of the public library \textsc{voro++} \citep{voropp}.

\item
\textit{Selection of candidates.}
A first selection of underdense regions is done by selecting all Voronoi cells that satisfy the criterion in local density contrast of $\delta_\mathrm{cell} < -0.4$.
Each underdense cell is considered the centre of a potential void.

\item
\textit{Growth of spheres.}
Centred on each candidate, the integrated density contrast $\Delta(r) = \delta(<r)$ is computed in spheres of increasing radius $r$ until it satisfies a redshift-dependent threshold of $\Delta_{\rm id} = -0.853$ for $\zsim=0.99$, obtained from the spherical evolution model \citep{sphcollapse1,sphcollapse2} by fixing a final spherical perturbation of $\Delta_\mathrm{id} = -0.9$ for $z=0$.

\item
\textit{Optimisation.}
Once these first void candidates are identified, step (iii) is repeated iteratively displacing the centre randomly a value proportional to $0.25$ times the previous radius until convergence to a sphere with maximum radius is achieved.
This procedure mimics a random walk around the original centre in order to obtain the largest possible sphere in that local minimum of the density field.

\item
\textit{Overlap filtering.}
Finally, the list of void candidates is cleaned so that each resulting sphere does not overlap with any other.
This cleaning is done by ordering the list of candidates by size and starting from the largest one.
The final result is a catalogue of non-overlapping spherical voids with a well-defined centre, a radius $R_\mathrm{v}$, and overall density contrast $\Delta(R_\mathrm{v}) = \Delta_\mathrm{id}$.

\end{enumerate}

The void finder also provides the position $\posv = (\possvx, \possvy, \possvz)$ $[\hmpc]$ and peculiar velocity $\velv = (\velvx, \velvy, \velvz)$ $[\kms]$ of the void centres.
Void velocities were computed summing all the individual velocities of haloes inside a spherical shell with dimensions $0.8 \leq r/R_\mathrm{v} \leq 1.2$.
This velocity is an unbiased and fair estimation of the bulk flow velocity of the void, as was demonstrated in \citet{lambas_sparkling_2016} (see their Fig.~1).

We do not consider AP distortions in this work.
In Section~\ref{subsec:impact_trsd} we provide a justification.
For this reason, we make use of the TC void catalogues of Paper~I, for which the same cosmology of the MXXL simulation was adopted in order to compute distances and densities, needed in void definition.
In order to study the impact of the RSD distortions, the identification was performed both in $r$-space and $z$-space.
In this context, $\rrs$ and $\rzs$ will denote void radius in both spatial configurations, respectively.

\subsection{Bijective mapping}
\label{subsec:data_map}

In Paper~I, we defined a bijective mapping between voids in $r$-space and $z$-space.
Specifically, for each $z$-space centre, we picked the nearest $r$-space centre with the condition that it must lay inside $1~\rzs$, removing all voids for which no partner could be found.
This mapping is a well defined function, since the condition of the nearest $r$-space neighbour assigns only one object to each $z$-space void.
This mapping is also injective, since the non-overlapping condition implies that each $r$-space void can only be reached by a single $z$-space object.
The filtering condition guarantees then a one-to-one relationship between voids in the two configurations.
These voids constitute what we call the \textit{bijective catalogues} (TC-rs-b and TC-zs-b in Table~1 from Paper~I, the former for the bijective voids in $r$-space, the latter for the bijective voids in $z$-space).
By construction, these catalogues have the same number of elements $(318,784)$, and it is ensured in this way that a void and its associated counterpart span the same region of space.
The original catalogues will be referred to as the \textit{full catalogues} (TC-rs-f and TC-zs-f in Table~1 from Paper~I) in order to distinguish them from the bijective ones.

In Fig.~\ref{fig:samples}, the left-hand panel shows the void abundances corresponding to the $z$-space catalogues, for both the full (grey dashed line) and bijective (grey solid line) versions.
The error bands were calculated from Poisson errors in the void counting process.
In Paper~I, we demonstrated that the void loss in the $z$-space mapping decreases as larger voids are considered.
Particularly, voids above the shot noise level are almost bijective.
A good indicator of this level is to take the median of the radius distribution.
In this case, it is $13.26~\hmpc$.
Visually, this can be appreciated in the figure by the small differences between the dashed and solid curves.
As a consequence, the full and bijective catalogues are statistically equivalent at these scales.
For this reason, we will not distinguish between them from now on, unless otherwise stated.

\subsection{Void sample}
\label{subsec:data_samples}

\begin{figure*}
    \includegraphics[width=\columnwidth]{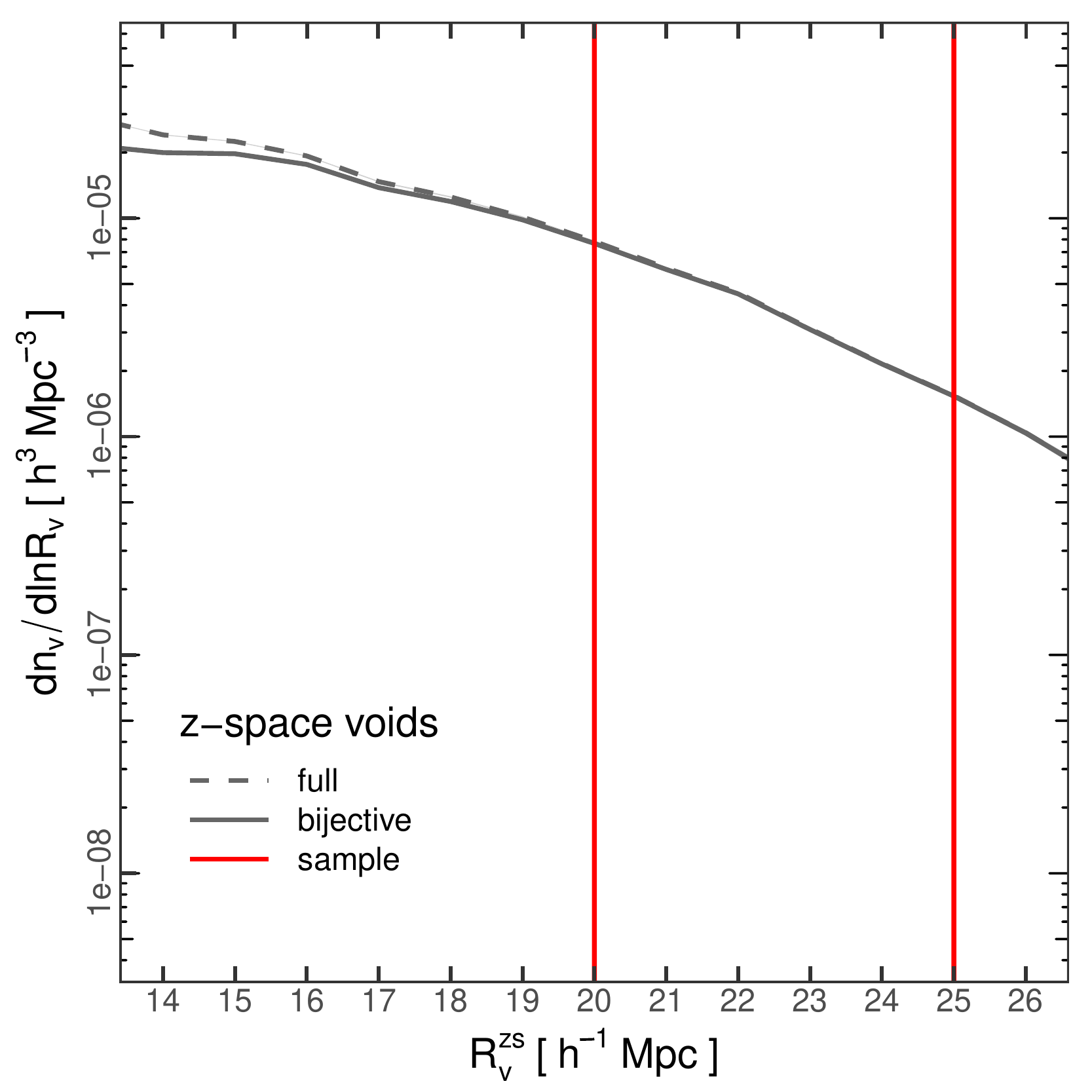}
    \includegraphics[width=\columnwidth]{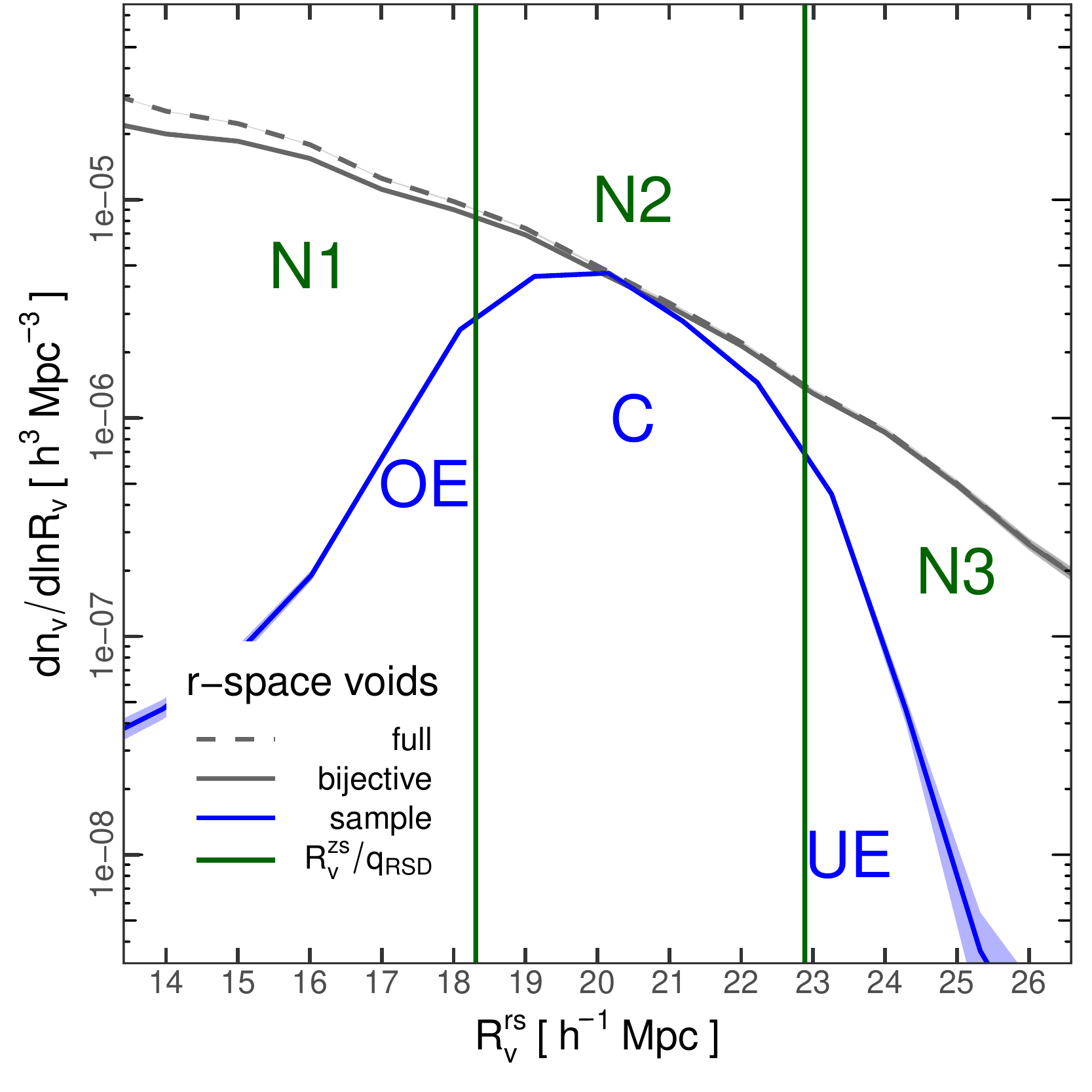}
    \caption{
    \textit{Left-hand panel.}
    Void abundances in redshift space corresponding to the full (grey dashed line) and bijective (grey solid line) catalogues.
    The void sample used to analyse the correlation function throughout the work is delimited by the red vertical lines.
    \textit{Right-hand panel.}
    The same abundances, but in real space.
    The blue solid curve describes how the real-space counterparts of the voids in the sample are distributed in radius.
    The green vertical lines represent the expansion effect correction of the sample by means of Eq.~(\ref{eq:qrsd}), a robust prediction of where the bulk of these voids must be found in real space.
    The subsamples built from these distributions, namely, N1, N2, N3, OE, C and UE, serve to understand the void ellipticity effect (see Section~\ref{subsec:impact_ellipticity}).
    }
    \label{fig:samples}
\end{figure*}

In order to measure the correlation function, we selected a sample of $z$-space voids with sizes between $20~\leq~\rzs/\hmpc~\leq~25$.
This cut is shown in Fig.~\ref{fig:samples} (left-hand panel) by the red vertical lines that delimit the sample.
We have verified that the general results do not depend on this cut, as long as it is carried out in the bijective range.

The right-hand panel of the figure shows the corresponding void abundances of the $r$-space catalogues.
In particular, the blue curve describes how the $r$-space counterparts of the voids in the sample are distributed in radius.
Note that, unlike the $z$-space voids, the $r$-space counterparts are not confined to a defined band, but they have a more complex distribution in $r$-space covering an extended range of scales, although notice that the y-axis is expressed in a logarithmic scale.
This is a central aspect in the analysis of this work, so we will come back to this figure later to explain the remaining features (see Section~\ref{sec:impact}).

\section{The projected correlation functions}
\label{sec:correlation}

We follow the methodology of C19 in order to measure correlations.
It is important to highlight that the original method relies on measuring correlations directly in terms of void-centric angles and redshifts between void-tracer pairs.
The cosmological dependence of these quantities with a physical distance scale is contained in the model, thus allowing us to test different cosmologies without the need to assume a fiducial one.
In this way, the AP effect is taken into account naturally.
However, as we are not considering AP distortions in this work, we treat correlations in the comoving coordinate system defined by the simulation and concentrate purely on RSD distortions.

Let us recall the basics here, in the context of these considerations.
$\xisp$ denotes the void-galaxy cross-correlation function in redshift space, where $\sigma$ represents the comoving void-centric distance in the plane of the sky (hereafter POS), and $\pi$ the analogue distance along the LOS.
Projecting $\xisp$ towards the POS in a given $\pi$-range, $\rm PR_\pi$, we get the \textit{POS correlation function}, $\xipos$, which is a function only of $\sigma$.
Conversely, projecting $\xisp$ towards the LOS in a given $\sigma$-range, $\rm PR_\sigma$, we get the \textit{LOS correlation function}, $\xilos$, a function only of $\pi$.

\subsection{Binning scheme}
\label{subsec:correlation_pr}

The measurement of the $z$-space correlation function is based on a cylindrical binning scheme.
In this geometry, a bin is a cylindrical shell oriented along the LOS, with internal radius $\sigma_{\rm int}$, external radius $\sigma_{\rm ext}$, a lower height $\pi_{\rm low}$, and an upper height $\pi_{\rm up}$.
For this study, in which the simulation box is taken as a simple mock survey, which is periodic, complete in volume and without any complicated selection functions, there is no need to employ estimators involving a random comparison sample (see for instance \citealt{estimator_landy}).
Therefore, the estimate of the correlation function can be considered analogous to that given by the natural estimator.
Given a bin labelled by $(i,j)$, and represented by the coordinates of its geometrical centre, $(\sigma_i, \pi_j)$, the estimator for the correlation function can be written as
\begin{equation}
    \hat{\xi}_{ij}(\sigma_i, \pi_j) = \frac{{\rm DD}_{ij}}{{\rm RR}_{ij}} - 1,
    \label{eq:estimator}
\end{equation}
where ${\rm DD}_{ij}$ represents the number of void-halo pairs counted inside the bin, and ${\rm RR}_{ij}$, the expected number of pairs in a homogeneous distribution, which in turn can be calculated analytically as the product of the density of haloes, the volume of the bin and the total number of voids.

The projected correlation functions can be considered as a special case in this binning scheme.
The POS projection is measured from a set of nested cylindrical shells distributed across the POS, whereas the LOS projection is measured from a string of filled cylinders oriented along the LOS.
More specifically, the scheme for the POS projection involves bins with dimensions $\sigma_{\rm int}$, $\sigma_{\rm ext}$, $\pi_{\rm low}=0$ and $\pi_{\rm up}={\rm PR_\pi}$, where $\delta \sigma = \sigma_{\rm ext} - \sigma_{\rm int}$ is the POS binning step.
Conversely, the scheme for the LOS projection involves bins with dimensions $\sigma_{\rm int}=0$, $\sigma_{\rm ext}={\rm PR_\sigma}$, $\pi_{\rm low}$ and $\pi_{\rm up}$, where $\delta \pi = \pi_{\rm up} - \pi_{\rm low}$ is the LOS binning step.
For the analysis of this work, we took equal projection ranges in both directions: ${\rm PR_\sigma} = {\rm PR_\pi} = 40 \hmpc$.
For simplicity, we refer to both quantities with the common notation ${\rm PR}$.
We also took equal binning steps: $\delta \sigma = \delta \pi = 1 \hmpc$.

\subsection{Configurations}
\label{subsec:correlation_config}

We measured correlations in different configurations of the spatial distribution of haloes and voids.
Measurements made with $z$-space voids and $z$-space haloes are referred to as the $\zxz$-space configuration.
Similarly, measurements made with $r$-space voids and $z$-space haloes are referred to as the hybrid $\rxz$-space configuration.
Finally, measurements made with $r$-space voids and $r$-space haloes are referred to as the $\rxr$-space configuration.
This notation also applies for measurements of the velocity field.

\begin{figure*}
    \includegraphics[width=\columnwidth]{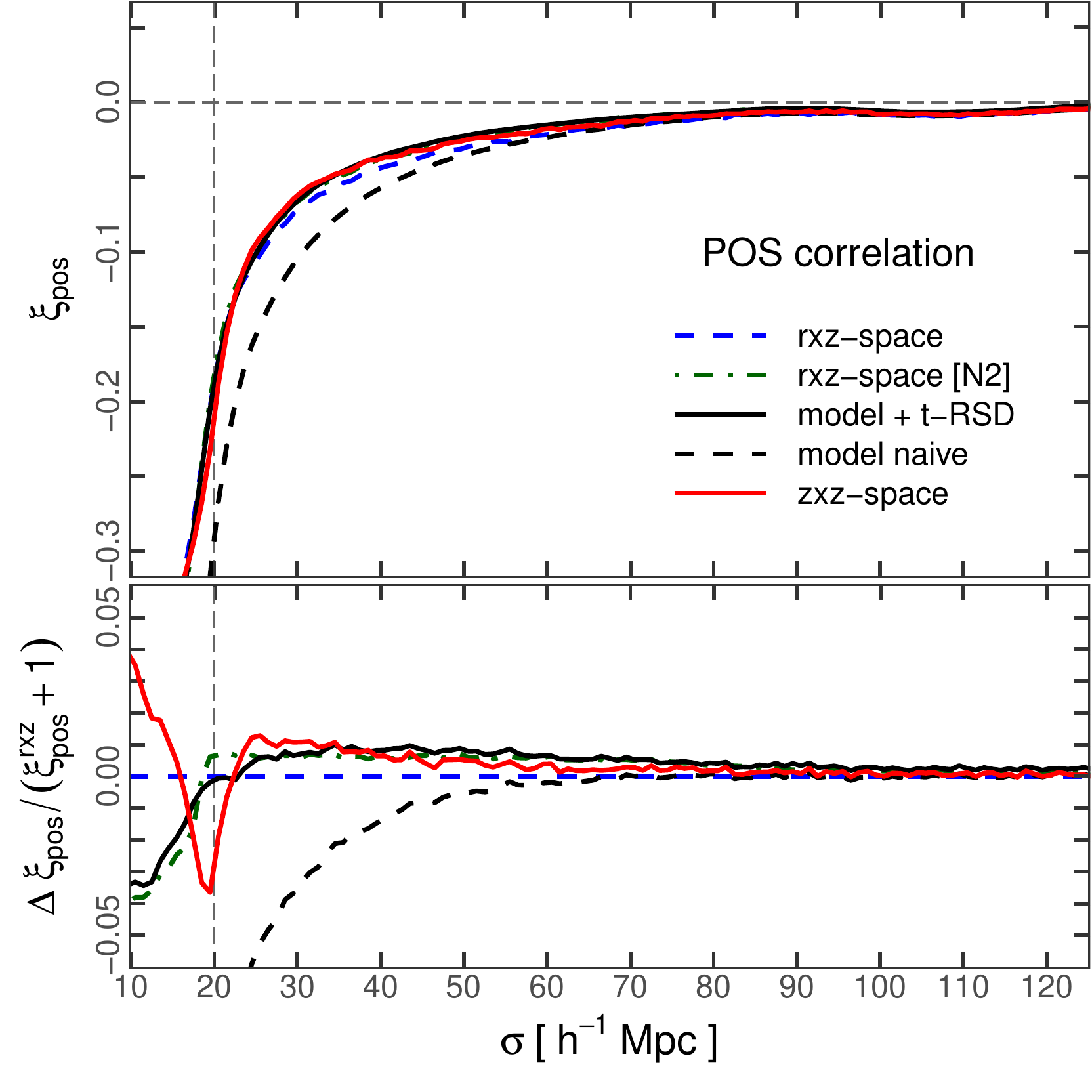}
    \includegraphics[width=\columnwidth]{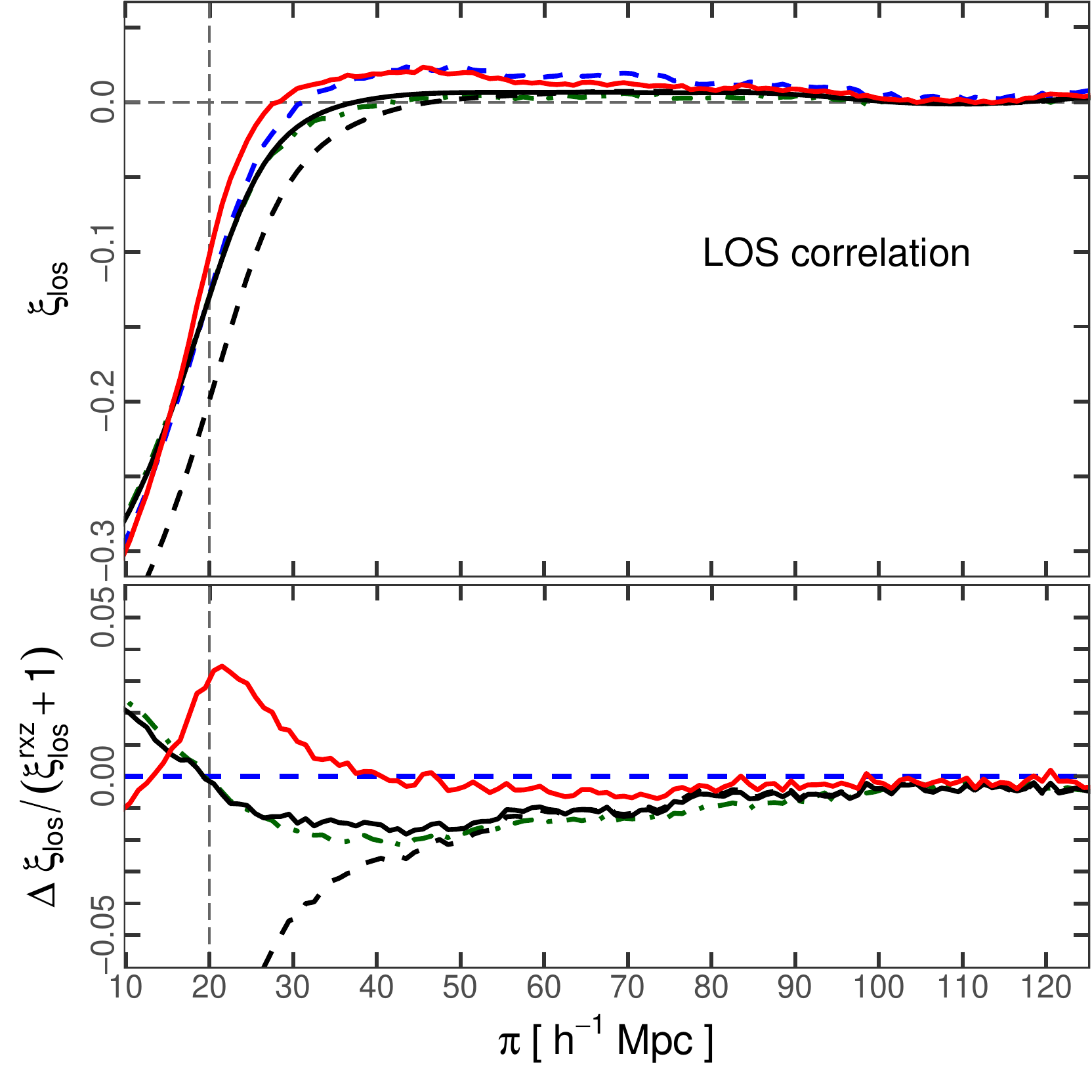}
    \caption{
    Projections of the correlation function towards the plane of the sky (left-hand panel) and towards the line of sight (right-hand panel).
    The red solid curves represent the measurements made in the $\zxz$-space configuration using the $z$-space sample defined in Section~\ref{subsec:data_samples} and shown in Fig.~\ref{fig:samples}.
    They represent a possible observational measurement.
    The blue dashed curves represent the measurements made in the hybrid $\rxz$-space configuration using the $r$-space counterpart of the sample.
    The black curves are two model predictions: without performing the expansion effect correction (naive, dashed lines) and after performing this correction (solid lines).
    The green dot-dashed curves represent the $\rxz$-space correlations using the sample N2, which allows us to test the performance of the tracer-RSD correction.
    In all cases, a projection range of $40~\hmpc$ was used.
    The lower panels show the corresponding fractional differences, taking the $\rxz$-space configuration as reference.
    }
    \label{fig:correlation}
\end{figure*}

Fig.~\ref{fig:correlation} shows the projected correlation functions corresponding to the void sample defined in Section~\ref{subsec:data_samples}.
The left-hand panel shows the POS projection, whereas the right-hand panel, the LOS projection.
Following C19, we are interested in scales greater than the minimum radius of the sample, in this case $20~\hmpc$.
The measurements made in the $\zxz$-space configuration are represented with a red solid line, i.e., these correlations are obtained from the spatial distribution of voids and haloes both in $z$-space, and hence, mimic a possible observational measurement.
The measurements made in the hybrid $\rxz$-space configuration, on the other hand, are represented with a blue dashed line, i.e., these correlations are obtained by taking the associated voids in $r$-space but keeping the haloes in $z$-space.
This is important because current models for RSD are defined to work in this hybrid configuration \citep{rsd_nadathur,reconstruction_nadathur}, and the aim of this paper is to compare model predictions with observations.
We will explain the meaning of the remaining curves in Section~\ref{sec:impact}.

It is more useful to compare correlations in different configurations by quantifying their fractional differences: $\Delta \xi / (\xi + 1) := (\xi_{\rm tar}-\xi_{\rm ref})/(\xi_{\rm ref}+1)$, where $\xi_{\rm tar}$ is the target correlation we want to compare, and $\xi_{\rm ref}$ is the one used as reference.
They are shown in the lower panels of the figure, where the hybrid $\rxz$-space configuration has been chosen as the reference one (blue dashed lines).

\subsection{Model}
\label{subsec:correlation_model}

In C19 we developed a physical model for the correlation function that takes into account the coupled distortions generated by the RSD and AP effects, along with the mixture of scales due to the geometry of the binning scheme.
We summarise its main aspects in this subsection.
For the following analysis we assume an arbitrary cylindrical binning scheme.
In this way, the model remains general and the projected correlation functions can be treated as a special case.
A central aspect to keep in mind is that this model is defined to work in the hybrid $\rxz$-space configuration.

The first step is to model the mixture of scales.
The correlation value corresponding to the bin $(i,j)$ can be predicted by expressing Eq.~(\ref{eq:estimator}) in differential form and then integrating over the volume of the bin:
\begin{equation}
    \xi_{ij}(\sigma_i, \pi_j) = 2 
    \frac{ \int_{\pi_{\rm low}}^{\pi_{\rm up}} d\pi \int_{\sigma_{\rm int}}^{\sigma_{\rm ext}} \sigma [1 + \xi(\sigma,\pi)] d\sigma }
    { (\sigma^2_{\rm ext}-\sigma^2_{\rm int}) (\pi_{\rm up}-\pi_{\rm low}) } - 1.
    \label{eq:scales}
\end{equation}
This equation was adapted from Eq.~(10) of C19 for the case of our simplified mock consisting of the simulation box and our hypothesis of working in a comoving coordinate system without AP distortions (instead of angles and redshifts as in the original method).

The next step is to provide a model for the correlation function taking into account the RSD effect.
Following \citet{xirsdmodel_peebles} and \citet{clues2}, $\xisp$ can be computed as the convolution of the $\rxr$-space correlation function, $\xi(r)$, and a pairwise velocity distribution of void-tracer pairs.
Note that $\xi(r)$ is a one-dimensional profile in view of the intrinsic spherical symmetry in $r$-space, with $r$ the distance to the centre of the voids.
This profile characterises the density contrast field around voids.
The pairwise velocity distribution can be assumed as a Gaussian distribution centered on the $\rxr$-space velocity profile describing the velocity field around voids, $v(r)$, with a constant velocity dispersion, $\sigma_v$.
We get the following expression\footnote{$\hat{\pi}$ refers to the irrational number \textit{pi}: 3.14159..., to avoid confusion with the $\pi$ coordinate.}:
\begin{equation}
    1 + \xi(\sigma, \pi) = 
    \int_{-\infty}^{\infty} [1 + \xi(r)] \frac{1}{\sqrt{2\hat{\pi}}\sigma_v}
    \mathrm{exp} \left[- \frac{(v_\parallel - v(r)\frac{r_\parallel}{r})^2}{2\sigma_v^2} \right]
    \mathrm{d}v_\parallel.
	\label{eq:gsm}
\end{equation}
This is the so-called \textit{Gaussian streaming model} (hereafter GS model).
Here, $r_\perp$ and $r_\parallel$ are the $\rxr$-space analogues of the $\rxz$-space quantities $\sigma$ and $\pi$.
Moreover, $v_\parallel$ denotes the LOS component of the void-centric peculiar velocity of tracers.
To complete the model, we need prescriptions for the $\rxr$-space velocity and density contrast profiles.

The velocity profile can be derived following linear theory via mass conservation up to linear order in density \citep{vel_peebles, clues2, aprsd_hamaus1, rsd_cai, rsd_nadathur}:
\begin{equation}
    v(r) = - \frac{1}{3} \frac{H(z)}{(1+z)} \beta(z) r \Delta(r).
	\label{eq:velocity}
\end{equation}
where $\Delta(r)$ is the integrated density contrast (introduced in Section~\ref{subsec:data_voids}), and $\beta(z)=f(z)/b$ is the so-called RSD parameter, i.e., the ratio between the logarithmic growth rate of density perturbations, $f(z)$, and the linear bias parameter, $b$, that relates the matter and halo density fields.
For the snapshot $z=0.99$ of the MXXL simulation that we are considering, $\beta=0.65$, obtained from the analysis carried out in C19.
In turn, $\Delta(r)$ can be explicitly related to $\xi(r)$ by the following expression:
\begin{equation}
	\Delta(r) = \frac{3}{r^3} \int_0^r \xi(r') r'^2 dr'.
    \label{eq:delta_int}
\end{equation}

\begin{figure*}
    \includegraphics[width=\columnwidth]{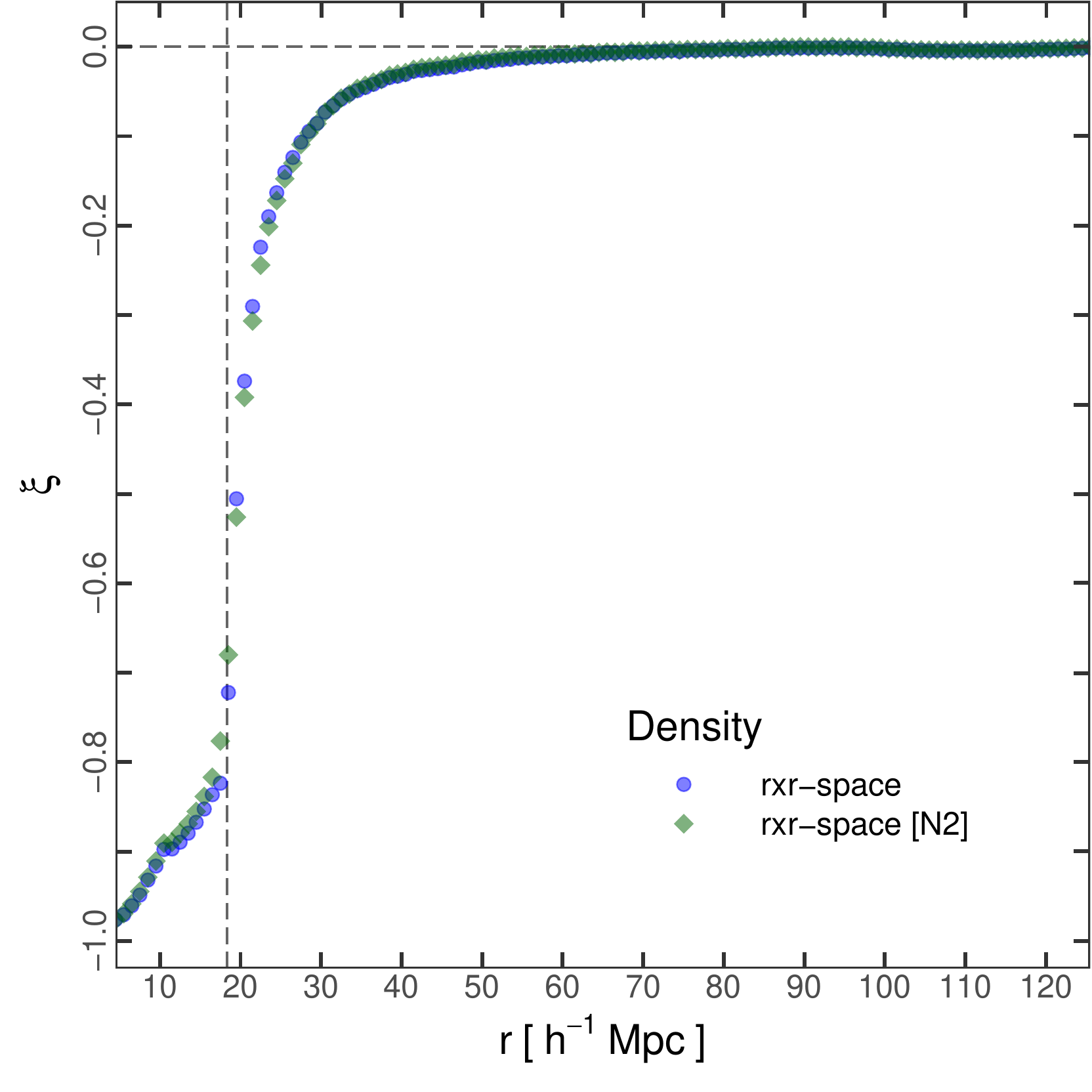}
    \includegraphics[width=\columnwidth]{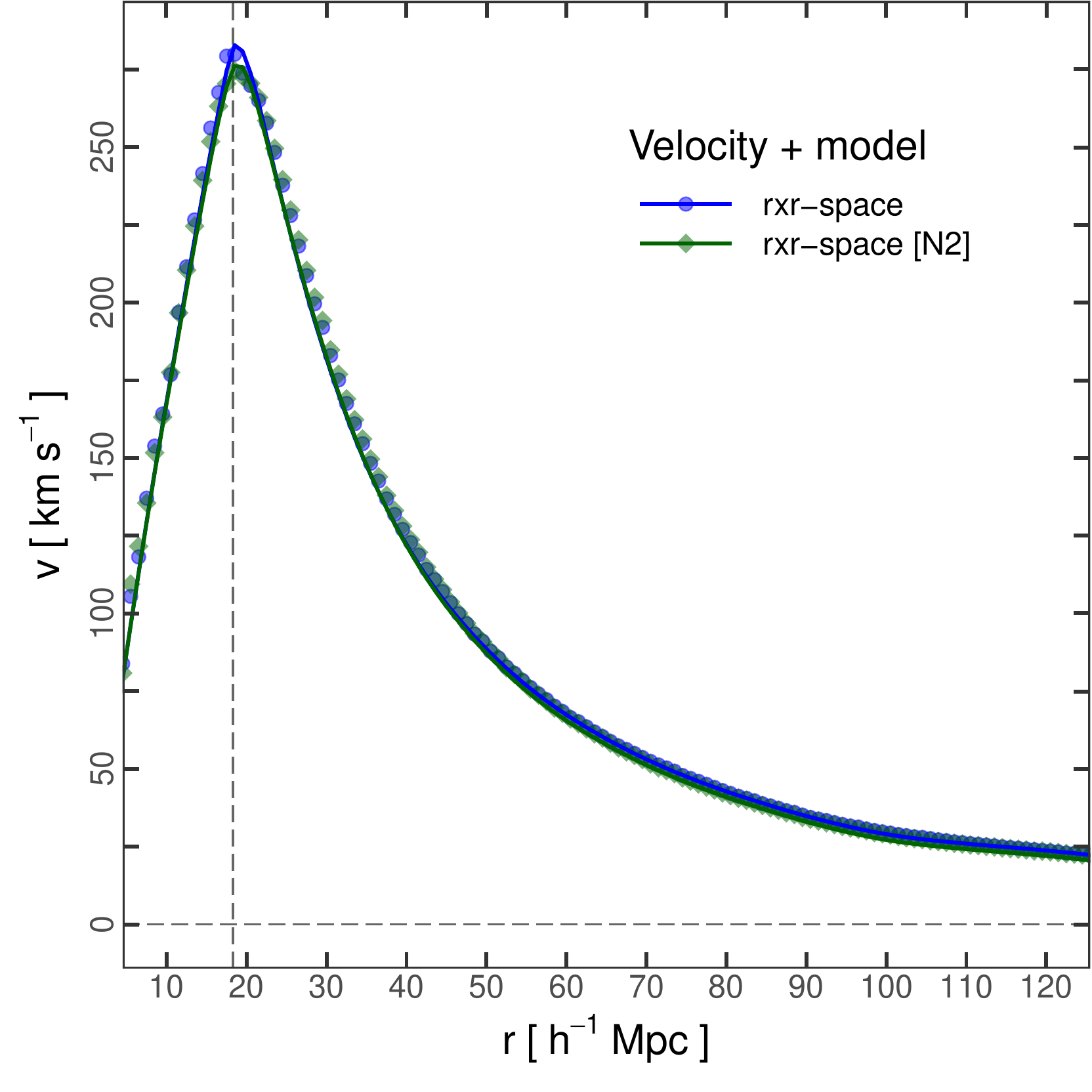}
    \caption{
    Void-centric density contrast (left-hand panel) and peculiar velocity (right-hand panel) profiles in $\rxr$-space.
    The blue dots correspond to the $r$-space counterpart of the original sample, whereas the green diamonds, to subsample N2.
    They are almost identical, which shows the potential of the tracer-RSD correction.
    The solid curves in the velocity analysis show the predictions of Eq.~(\ref{eq:velocity}), which works remarkably well at all scales.
    } 
    \label{fig:profiles}
\end{figure*}

Unlike $v(r)$, there is not a successful model for the density profile derived from first principles, but it is a common practice to use parametric and empirical approaches (see for instance \citealt{clues2, density_hamaus, density_nadathur}).
In particular, we provided one in C19 (see Eq.~15).
Nevertheless, we do not use any of them in this work.
Instead, we use the $\rxr$-space profile of the sample directly measured from the simulation as input in the model.
This allows us to understand with precision all the $z$-space systematics that affect the correlation function, the aim of this paper, since we are not introducing additional effects due to the performance of the density models.
This profile is shown at the left-hand panel of Fig.~\ref{fig:profiles} with blue dots.
The right-hand panel shows the corresponding velocity profile, also with blue dots.
The blue solid line is the prediction of Eq.~(\ref{eq:velocity}), which works remarkably well at all scales.
We will explain the meaning of the remaining profiles displayed in the figure (represented with green diamonds) in Section~\ref{subsec:impact_trsd}.

\section{Redshift-space effects in voids}
\label{sec:zeffects}

In Paper~I, we carried out a thorough investigation of the $z$-space effects that affect void properties, namely, their number, size and spatial distribution.
Particularly, we focused on the impact that these effects have on the void size function taken as a cosmological test.
We continue the analysis in this paper, focusing now on the void-galaxy cross-correlation function.
Before that, in this section, we review the main aspects of the physical description of these effects with the aim to provide the framework to analyse the anisotropic patterns observed on the correlation measurements, and in this way, to figure out all the sources that cause them.
This analysis is fundamental to obtain unbiased cosmological constraints.
But besides this practical importance, it will also shed light on important features of the nature of cosmic voids regarding their structure and dynamics.

The pillar of the analysis resides in the one-to-one relationship between $z$-space and $r$-space voids above the shot-noise level.
Incidentally, this means that the conservation of void-tracer pairs is a good assumption when modelling RSD around voids.
In view of this, any systematic difference in the statistical properties between both void populations can only be attributed to some physical transformation that voids suffer when they are mapped from $r$-space into $z$-space, and hence, such transformation must be associated to the distortions in the observed spatial distribution of the galaxies.
Consequently, a physical description of the phenomenon must find its bases on the large-scale dynamics and from cosmological considerations.

\subsection{Expansion effect}
\label{subsec:zeffects_trsd}

In Paper~I, we showed that $z$-space voids are systematically bigger than their $r$-space counterparts.
The reason is that voids expand when they are mapped from $r$-space into $z$-space.
This is the \textit{expansion effect}, a consequence of the RSD induced by tracer dynamics at scales around the void radius.
We refer to this effect, particularly to the type of distortions that it generates, with the acronym tracer-RSD (or simply t-RSD).

A description of this effect is possible from the basis of Eq.~(\ref{eq:velocity}), from which a linear relation is found between both radii, $\rrs$ and $\rzs$:
\begin{equation}
    \rzs = \qrsd \rrs, \\
    \qrsd = 1 - \frac{1}{3} \delta R_{\rm v} \beta(\zsim) \Delta_{\rm id}.
    \label{eq:qrsd}
\end{equation}
Here, $\delta R_{\rm v}$ is a parameter that quantifies variation in radius, and $\Delta_{\rm id}$ is the threshold value for void identification introduced in Section~\ref{subsec:data_voids}.
In Paper~I, we showed that voids above the shot noise level favor the value $\delta R_{\rm v}=0.5$.
Then, the explicit value of the proportionality factor corresponding to the MXXL simulation is $\qrsd = 1.092$.
Note that $\qrsd > 1$, hence $\rzs > \rrs$, in agreement with our assumption that voids expand in $z$-space.

\subsection{Off-centring effect}
\label{subsec:zeffects_vrsd}

In Paper~I, we showed that $z$-space void centres are systematically shifted along the LOS.
This is the \textit{off-centring effect}, a consequence of a different class of RSD induced by large scale flows in the matter distribution.
Interpreting voids as whole entities that move in space with a net velocity \citep{lambas_sparkling_2016,ceccarelli_sparkling_2016,lares_sparkling_2017}, this effect can be considered as a by-product of the RSD induced by void dynamics.
We refer to this effect, particularly to the type of distortions that it generates, with the acronym void-RSD (or simply v-RSD).

An analytical prediction of this effect can be obtained from an expression equivalent to Eq.~(\ref{eq:halo_zspace}) for the case of voids:
\begin{equation}
    \Tilde{X}_{\rm v 3} = \possvz + \velvz \frac{(1 + z)}{H(z)},
	\label{eq:vrsd}
\end{equation}
where $\Tilde{X}_{\rm v3}$ denotes the shifted $\possvz$-coordinate.
In Paper~I, we also showed that, although the tracer-RSD and void-RSD effects manifest together in observations, they can be considered independent with each other.

\subsection{Alcock-Paczynski volume effect}
\label{subsec:zeffects_ap}

The AP effect introduces additional distortion patterns in the spatial distribution of the galaxies surrounding voids.
As a consequence, the volume of voids is also altered by this effect.
This is because the discrepancies between the fiducial and true cosmologies generate deviations between the POS and LOS dimensions of voids, which must be spherical in $r$-space.
This is the \textit{Alcock-Paczynski volume effect}.

Following a similar approach to that used for the expansion effect, it is possible to relate the true and fiducial radii, $\rrs$ and $R_{\rm v}^{\rm fid}$:
\begin{equation}
    R_{\rm v}^{\rm fid} = \qap \rrs, \\
    \qap = \sqrt[3]{(\qap^\perp)^2 \qap^\parallel},
	\label{eq:qap}
\end{equation}
where
\begin{equation}
    \qap^\perp = \frac{D_{\rm M}^{\rm fid}(z)}{D_{\rm M}^{\rm true}(z)}, \\
    \qap^\parallel = \frac{H_{\rm true}(z)}{H_{\rm fid}(z)}.
	\label{eq:qap_pos_los}
\end{equation}
Here, $D_{\rm M}$ denotes the comoving angular diameter distance.
The indices "fid" and "true" refer to fiducial and true quantities, respectively.
Unlike the expansion effect, the net result of the AP volume effect is not necessarily an expansion, it can also be a contraction.
This is because $\qap$ can be less or greater than one, depending on the fiducial cosmological parameters chosen.

Note that the RSD and AP factors are simply two constants of proportionality, independent of the scale, and strongly dependent on the cosmological parameters.
Furthermore, there is an interesting difference between them.
On the one hand, $\qap$ depends only on the background cosmological parameters, such as $\Omega_m$ and $H_0$, hence it is related to the expansion history and geometry of the Universe.
On the other hand, $\qrsd$ depends only on $\beta$, so it is a dynamical parameter related to the growth rate of cosmic structures.

Actually, the volume of voids is affected by the combined contribution of the tracer-RSD and AP effects.
In Paper~I, we showed that both of them can be considered as independent effects, hence, void radii can be related by combining Eqs.~(\ref{eq:qrsd}) and (\ref{eq:qap}), resulting in the following two-step correction:
\begin{equation}
    \rzs = \qap ~ \qrsd ~ \rrs.
    \label{eq:q_ap_rsd}
\end{equation}
This is all it takes to correct an observational void size function.

\section{Impact on the correlation function}
\label{sec:impact}

The goal of this section is to understand all the factors that contribute to the final anisotropic shape of the projected correlation functions measured in the $\zxz$-space configuration, the red solid curves of Fig.~\ref{fig:correlation}, which represent a possible observational measurement from a survey.
This analysis is based on the framework developed on the previous section.
Although we analyse the projected version of the correlation function, the main conclusions we reach are general.
Moreover, we have checked that the following results are qualitatively independent of the projection range.
Here, we show the case $\mathrm{PR} = 40~\hmpc$, as this constitutes a realistic case applicable to data (see C19 for more details).

\subsection{Impact of the impurity of a sample}
\label{subsec:impact_impurity}

In Section~\ref{subsec:data_map}, we mentioned that the full and bijective void catalogues are statistically equivalent at the scales of interest, hence we were not going to distinguish between them.
We make an exception here to reinforce this concept concerning the correlation function.

\begin{figure}
    \includegraphics[width=\columnwidth]{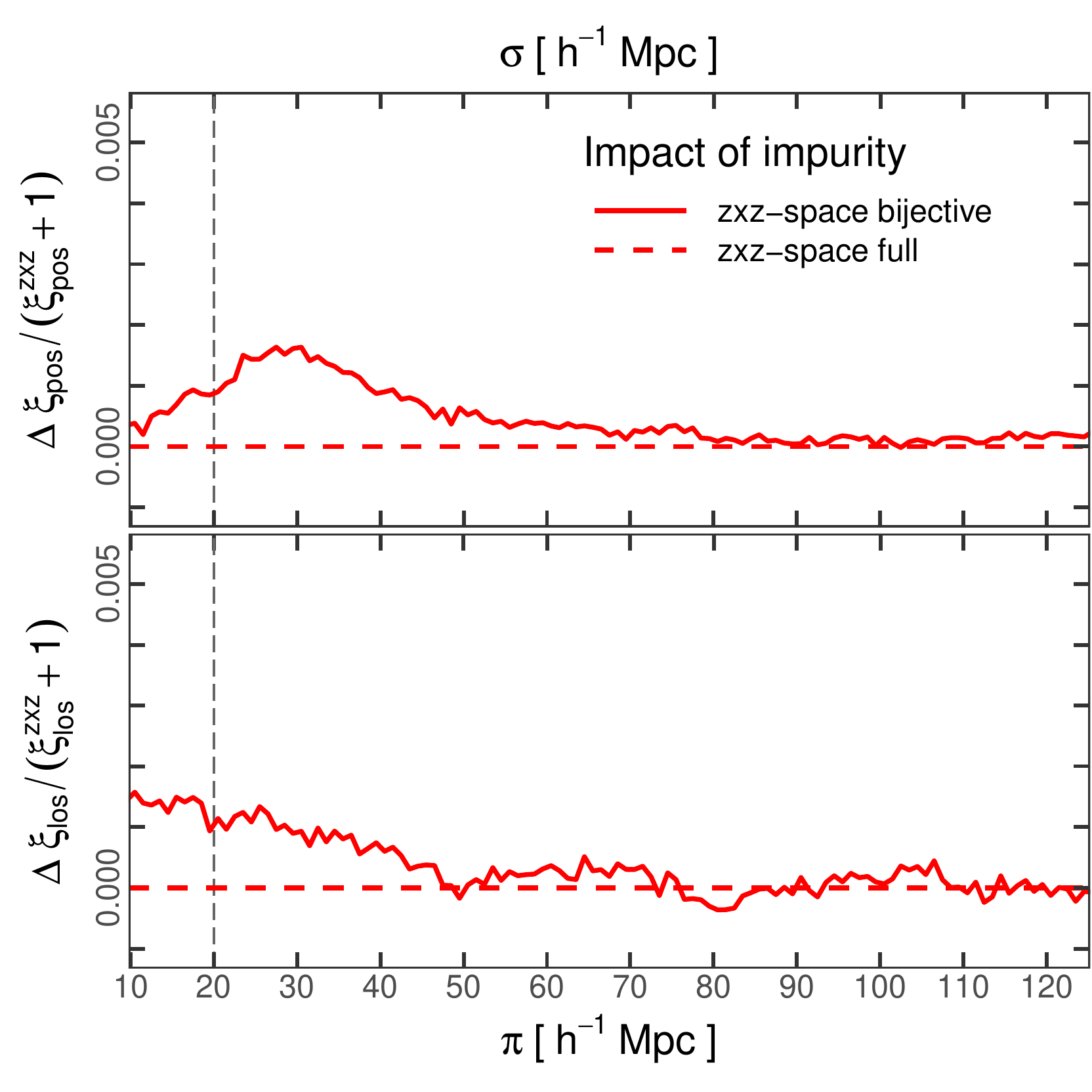}
    \caption{
    Impact of the impurity of a sample regarding non-bijective voids quantified by means of the fractional differences of correlation between the measurements made with the original sample, obtained from the bijective catalogue, and with an analogous sample taken from the full catalogue.
    The differences are very small, less than $0.2\%$, allowing us to treat both samples as statistically equivalent. 
    }
    \label{fig:impurity}
\end{figure}

In Section~\ref{subsec:data_samples}, we selected the void sample to use throughout the work.
It is composed of $z$-space voids in the range $20~\leq~\rzs/\hmpc~\leq~25$.
Technically, this cut was applied to the bijective catalogue.
We repeated the analysis applying the same cut, but this time to the full catalogue.
Fig.~\ref{fig:impurity} shows the fractional differences between the correlations measured with these two samples.
The upper panel corresponds to the differences between the POS projections, whereas the lower panel, to the LOS projections.
Note that the scale was augmented with respect to the scale shown in Fig.~\ref{fig:correlation} in order to highlight the magnitude of the differences. 
They are very small, less than $0.2\%$ at all scales.
In view of this, the impurity of a sample regarding non-bijective voids produces a negligible impact on correlation measurements.

\subsection{Impact of the expansion effect}
\label{subsec:impact_trsd}

We can now make a first model prediction.
As was explained in Section~\ref{subsec:correlation_model}, the model needs the $\rxr$-space density profile, $\xi(r)$, as input.
As a first and naive ansatz, one could imagine that the correct profile is that corresponding to the $r$-space voids with radii in the same range as the $z$-space sample: $20~\leq~\rrs/\hmpc~\leq~25$.
The model prediction using this profile is shown in Fig.~\ref{fig:correlation} with a black dashed line.
It is clear that it completely fails to reproduce the $\zxz$-space correlations: compare the black dashed curves with the red solid ones.
This failure of the model was expected, since it is defined to operate in the hybrid $\rxz$-space configuration, as it does not take into account the global dynamics of voids.
However, it also fails to reproduce the measurements in this configuration too, as evident in the figure comparing now the black dashed curves with the blue dashed ones.
The relative differences can be as high as $8\%$, specially near the void walls.

The inefficiency of the model highlights the importance of relating to the correct $\rxr$-space statistical properties of the sample.
The mistake is to assume that the $r$-space counterparts of the $z$-space voids in the sample fall in the same range of the radius distribution.
The right-hand panel of Fig.~\ref{fig:samples} shows that the corresponding $r$-space distribution is more complex, covering an extended range of scales.
In view of this, we repeated the analysis using exactly all the $r$-space counterparts of the $z$-space voids in the sample to measure the $\rxr$-space density profile.
This is the profile that we have shown in Fig.~\ref{fig:profiles} (left-hand panel) with blue dots.
The corresponding model prediction for the projected correlations is shown in Fig.~\ref{fig:correlation} with a black solid line.
Note that the deviations have been noticeably reduced, although not completely: compare the black solid curves with the blue dashed ones.
There is a remaining deviation of order $2\%$ at scales near the void walls.

The immediate question that arises now is how to describe this complex $r$-space sample.
The answer lies in Eq.~(\ref{eq:qrsd}), namely, in the expansion effect.
According to this effect, the $r$-space sample should fall in the range $20/\qrsd~\leq~\rrs/\hmpc~\leq~25/\qrsd$.
This is shown at the right-hand panel of Fig.\ref{fig:samples} with green vertical lines.
Note that the range thus delimited captures the bulk of the voids that constitute the sample.
Let us call all the $r$-space voids inside this band with the name N2, for reasons that will be clarified in Section~\ref{subsec:impact_ellipticity}.
The sample N2 approximates the true $r$-space sample with respect to its statistical properties.
This is evident in Fig.~\ref{fig:profiles}, where we show the density and velocity profiles of the sample N2 with green diamonds.
Note that it is almost identical to the profiles corresponding to the true $r$-space sample (blue dots).
This demonstrates the ability of the tracer-RSD correction to recover the $\rxr$-space statistical properties of voids identified from observational data.

Up to here, we can assert that the most important source of distortion patterns on the correlation function is the tracer-RSD expansion effect, which can be accounted for by the GS model combined with Eq.~(\ref{eq:qrsd}).
The remaining deviations between the observations (red solid curves) and the model prediction (black solid curves) can be separated into two components.
The hybrid $\rxz$-space configuration (blue dashed curves) will serve as a mediator between both of them.
We explore this in the following subsections.

Before moving on, a brief comment about the impact of the AP volume effect on the correlation function.
Although we are not considering this effect in this work, it is completely analogous to the expansion effect.
Concretely, the radius distribution behaves similarly to Fig.~\ref{fig:samples}.
The only difference is that the bulk of $r$-space voids are found now using Eq.~(\ref{eq:q_ap_rsd}) instead of Eq.~(\ref{eq:qrsd}), which combines the contribution of both the tracer-RSD and AP effects.
More specifically, they are found in the range $20/(\qap~\qrsd)~\leq~\rrs/\hmpc~\leq~25/(\qap~\qrsd)$.

\subsection{Impact of the off-centring effect}
\label{subsec:impact_vrsd}

The first component of the remaining deviations is related to the differences between the correlations in the $\zxz$- and $\rxz$-space configurations (red solid and blue dashed curves in Fig.~\ref{fig:correlation}).
In the context of the bijective mapping between voids, these differences can only be attributed to the displacements of centres when they are mapped from $r$-space into $z$-space, i.e. to the off-centring effect.
We highlight the fact that this effect is responsible for an additional distortion pattern that is different from the classic RSD due to tracer dynamics (tracer-RSD), but it is due to void dynamics (void-RSD).
It is also true that these deviations are smaller than those corresponding to the expansion effect, being at most of the order $3\%$ at scales near the void walls.

\begin{figure}
    \centering
    \includegraphics[width=\columnwidth]{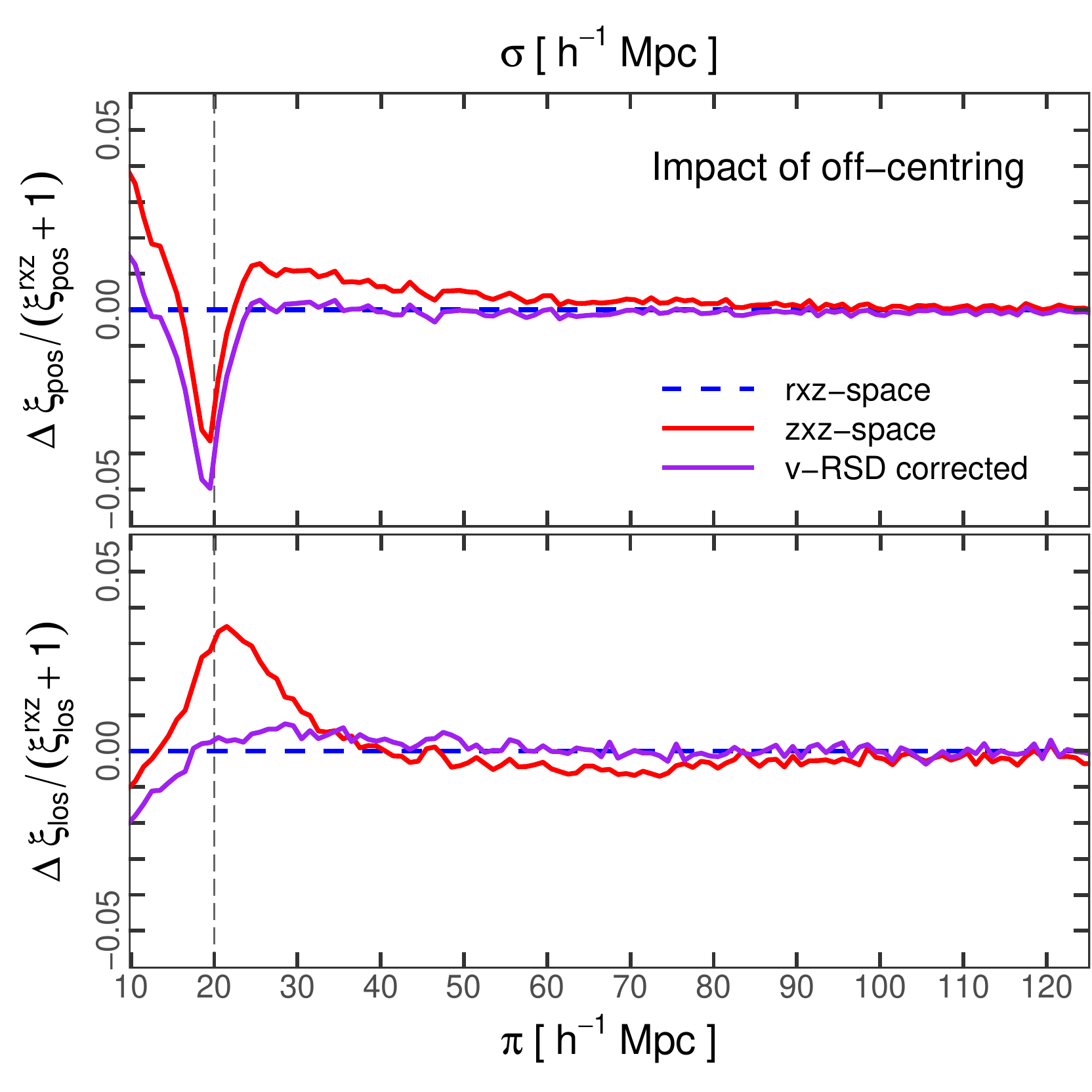}
    \caption{
    Impact of the off-centring effect quantified by means of the fractional differences of correlation between the measurements made in the $\zxz$-space (red solid lines) and $\rxz$-space (blue dashed lines) configurations.
    In the context of the bijective mapping, these differences can only be attributed to the displacements of void centres under this mapping.
    They represent an additional distortion pattern not previously taken into account, which is due to the global dynamics of the regions containing the voids (void-RSD).
    Note that these deviations diminish notably after correcting the position of the $z$-space centres with Eq.~(\ref{eq:vrsd}) (purple solid lines).
    }
    \label{fig:vrsd}
\end{figure}

We can quantify physically the void-RSD distortions using Eq.~(\ref{eq:vrsd}).
Fig.~\ref{fig:vrsd} shows what happens if the $z$-space centre positions are corrected with this equation (purple solid lines).
The deviations diminish notably, with differences well bellow $1\%$, thus recovering the correlation function in the hybrid $\rxz$-space configuration remarkably well at all scales of interest.
This is the first time that this type of distortions on the correlation function is detected and quantified.

The correction performed here was possible because we are working with a simulation.
The void finder can compute the velocity of voids from the individual velocity of tracers, hence Eq.~(\ref{eq:vrsd}) can be applied to correct the centres.
Nevertheless, this is not possible in practice.
One feasible solution is to incorporate a void velocity distribution (see Fig.~6 of Paper~I and its description) into the GS model.
We leave this topic for a future investigation.

\subsection{Impact of void ellipticity}
\label{subsec:impact_ellipticity}

The second component of the remaining deviations is related to the differences between the measurements in the $\rxz$-space configuration and the model prediction (blue dashed and black solid curves in Fig.~\ref{fig:correlation}), which are of order $2\%$ at scales near the void walls, as was noted in Section~\ref{subsec:impact_trsd}.

Given the fact that the sample N2 (obtained with the tracer-RSD correction) approximates well the true $r$-space sample, it is reasonable to expect that the correlation functions measured with both of them (in the $\rxz$-space configuration) will be almost identical.
However, this does not happen.
In Fig.~\ref{fig:correlation}, the projected correlations corresponding to the sample N2 are shown with a green dot-dashed line.
A comparison with the blue dashed curves shows that there are appreciable deviations, specially in the LOS projection.
This is a completely unexpected result, since both samples have almost identical $\rxr$-space density and velocity profiles, as we showed in the analysis of Fig.~\ref{fig:profiles}.
Evidently, the tails of the $r$-space sample distribution have an appreciable effect on correlations. 
Note however, that the model predicts remarkably well the N2 correlations, finding negligible differences between them: compare the black solid curves with the green dot-dashed ones.
This result ratifies the methodology developed in C19.

To find an explanation, we need to go back to the right-hand panel of Fig.~\ref{fig:samples} and define some subsamples of voids.
We have already defined one of them: the subsample N2, with all voids inside the band delimited by the tracer-RSD correction with the factor $\qrsd$ (green vertical lines).
Analogously, the subset of voids that also fall inside this band, but under the distribution of the overall $r$-space sample (blue solid curve), will constitute the subsample C.
Note that these two subsamples are very similar in view of the tracer-RSD correction.
Now, the voids from the left tail will constitute the subsample OE, whereas the voids from the right tail, the subsample UE.
In the same manner, all voids that fall at the left of the band, but in the same range as subsample OE, will constitute the subsample N1.
Similarly, all voids at the right of the band, but in the same range as subsample UE, will constitute the subsample N3.
Note that the subsamples N1, N2 and N3 contain the subsamples OE, C and UE, respectively.

\begin{figure*}
    \includegraphics[width=175mm]{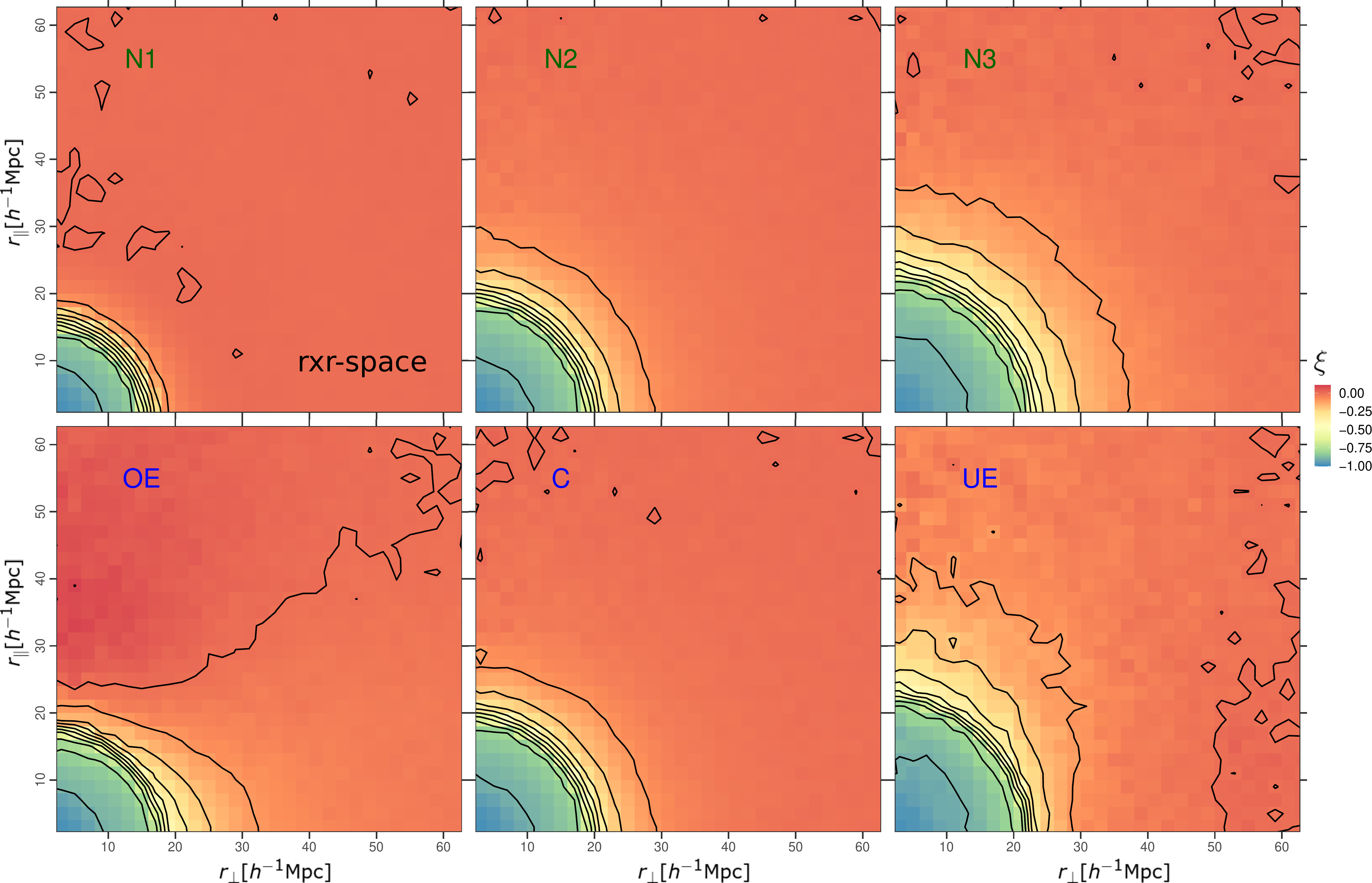}
    \caption{
    Density contrast maps in $\rxr$-space corresponding to the subsamples defined in Fig.~\ref{fig:samples}.
    The normal subsamples (upper panels) exhibit circular contours with no signs of anisotropies, as expected.
    The only difference between them is in the size of the empty regions, in agreement with the fact that, from N1 to N3, they are composed of voids of increasing radii.
    Subsample C (lower middle panel), in addition, behaves similarly to subsample N2.
    This highlights the power of the tracer-RSD correction.
    The remaining subsamples, OE and UE, by contrast, exhibit prominent and opposite anisotropic patterns.
    This is a manifestation of the ellipsoidal nature of voids.
    } 
    \label{fig:map2d_dens}
\end{figure*}

Fig.~\ref{fig:map2d_dens} shows the two-dimensional $\rxr$-space density field, $\xi(r_\perp,r_\parallel)$, for each of the subsamples defined above.
On each map, the blue areas describe the emptiest regions.
All maps were colour-coded using the same scale.
The first relevant feature to be outlined is that the subsamples N1, N2 and N3 (upper panels) behave as expected: they exhibit circular contours with no signs of anisotropies.
The only difference among them is in the size of the empty regions, in agreement with the fact that, from N1 to N3, they are composed of voids of increasing radii.
For this reason, these subsamples have been labelled with the letter N, accounting for \textit{normal}.
Note that the subsample C (lower middle panel) also behaves normally, and in particular, very similarly to subsample N2. 
Note however, that the subsamples OE and UE (lower left and lower right-hand panels) behave unexpectedly.
They exhibit prominent anisotropic patterns.
This is an interesting result, since we are analysing the density field in $\rxr$-space, and hence, expecting spherical symmetry.
Furthermore, the anisotropies are opposite: OE voids are elongated along the POS axis, whereas UE voids are elongated along the LOS axis.

\begin{figure*}
    \includegraphics[width=175mm]{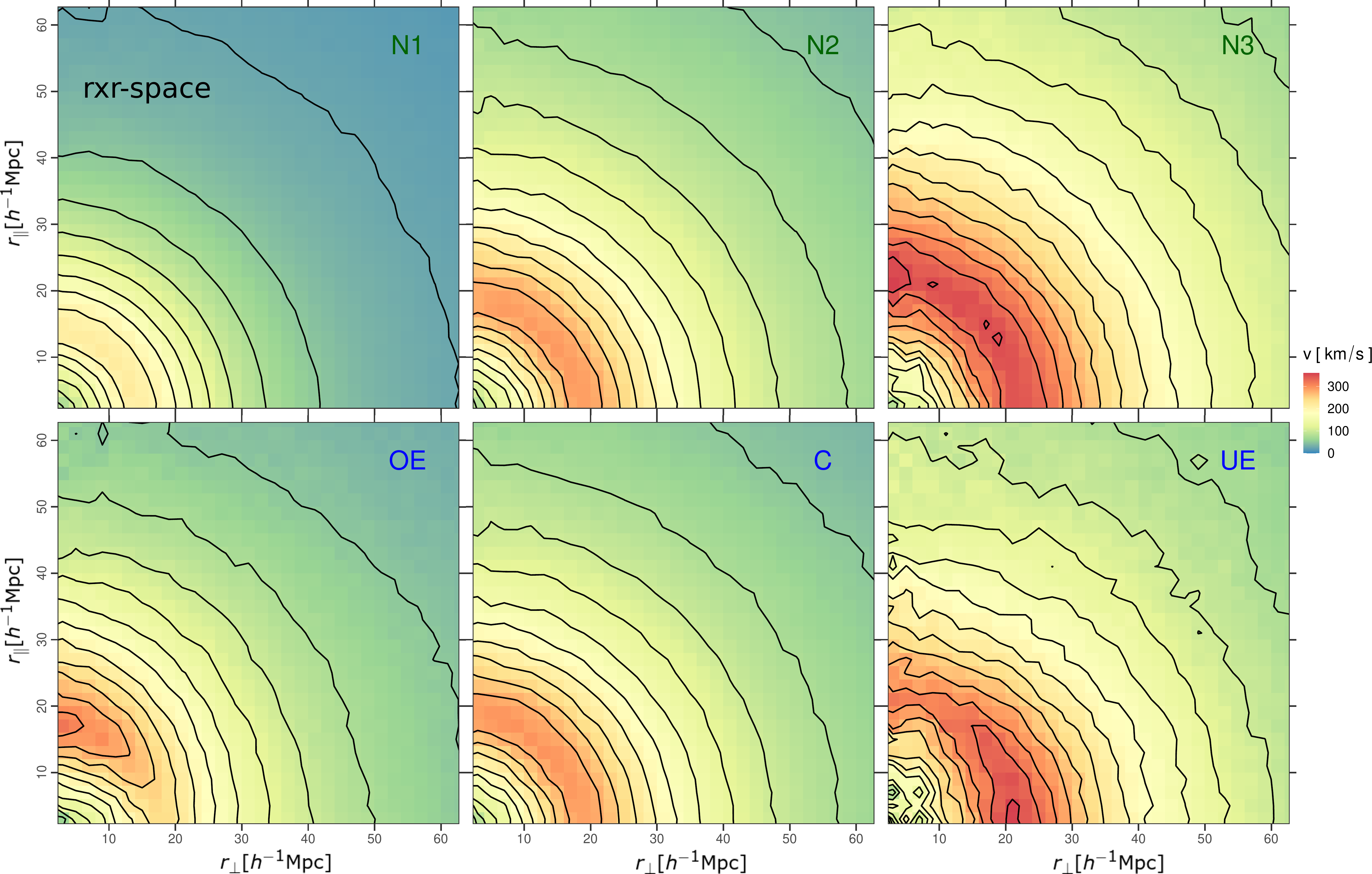}
    \caption{
    Velocity maps in $\rxr$-space as a complementary information of Fig.~\ref{fig:map2d_dens}.
    The same behavior is observed.
    Voids from subsample OE are over-expanding voids, since their $z$-space radii are greater than the prediction of Eq.~(\ref{eq:qrsd}), whereas those from subsample UE are under-expanding voids, since their radii are smaller.
    }
    \label{fig:map2d_vel}
\end{figure*}

Fig.~\ref{fig:map2d_vel} shows the associated $\rxr$-space velocity field, $v(r_\perp,r_\parallel)$, for each of the subsamples.
As in the case of the density, each map was colour-coded using the same scale.
The same results are found: the subsamples N1, N2 and N3 exhibit circular contours, the subsample C behaves similarly to subsample N2, whereas the subsamples OE and UE exhibit prominent and opposite anisotropic patterns.

We interpret these results in the following way.
Individual voids are typically ellipsoidal, but oriented randomly in space.
Therefore, this ellipticity has no significant impact on the statistical properties of a stacking of a complete sample of voids.
This is the case of the subsamples N1, N2 and N3.
This is also the case of the subsample C, since it is very similar to subsample N2.
This is a feature of the tracer-RSD correction: it predicts a completeness region in the void radius distribution.
For this reason, this subsample has been labelled with the letter C, accounting for \textit{complete}.
Voids from the subsamples OE and UE, however, are not complete, but constitute a special selection of voids.
They do not follow the velocity and expansion predictions of Eqs.~(\ref{eq:velocity}) and (\ref{eq:qrsd}), respectively.
On the one hand, OE voids are \textit{over-expanding voids}, since they fall in the selection range when they are identified in $z$-space, but their radii are greater than the prediction given by the factor $\qrsd$.
Conversely, UE voids are \textit{under-expanding voids}, since they also fall in the selection range, but their radii are lower than the corresponding prediction.

\begin{figure*}
    \includegraphics[width=\columnwidth]{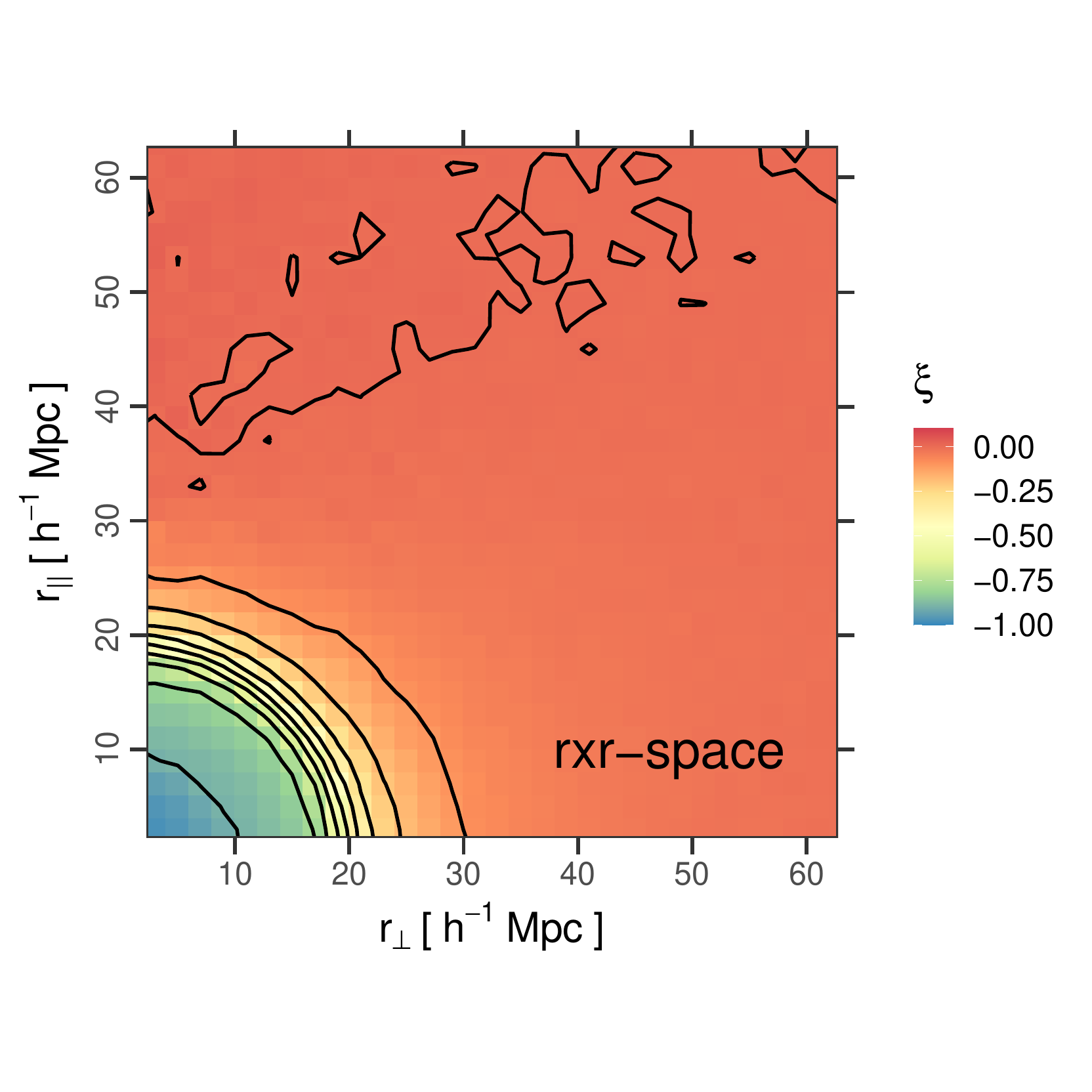}
    \includegraphics[width=\columnwidth]{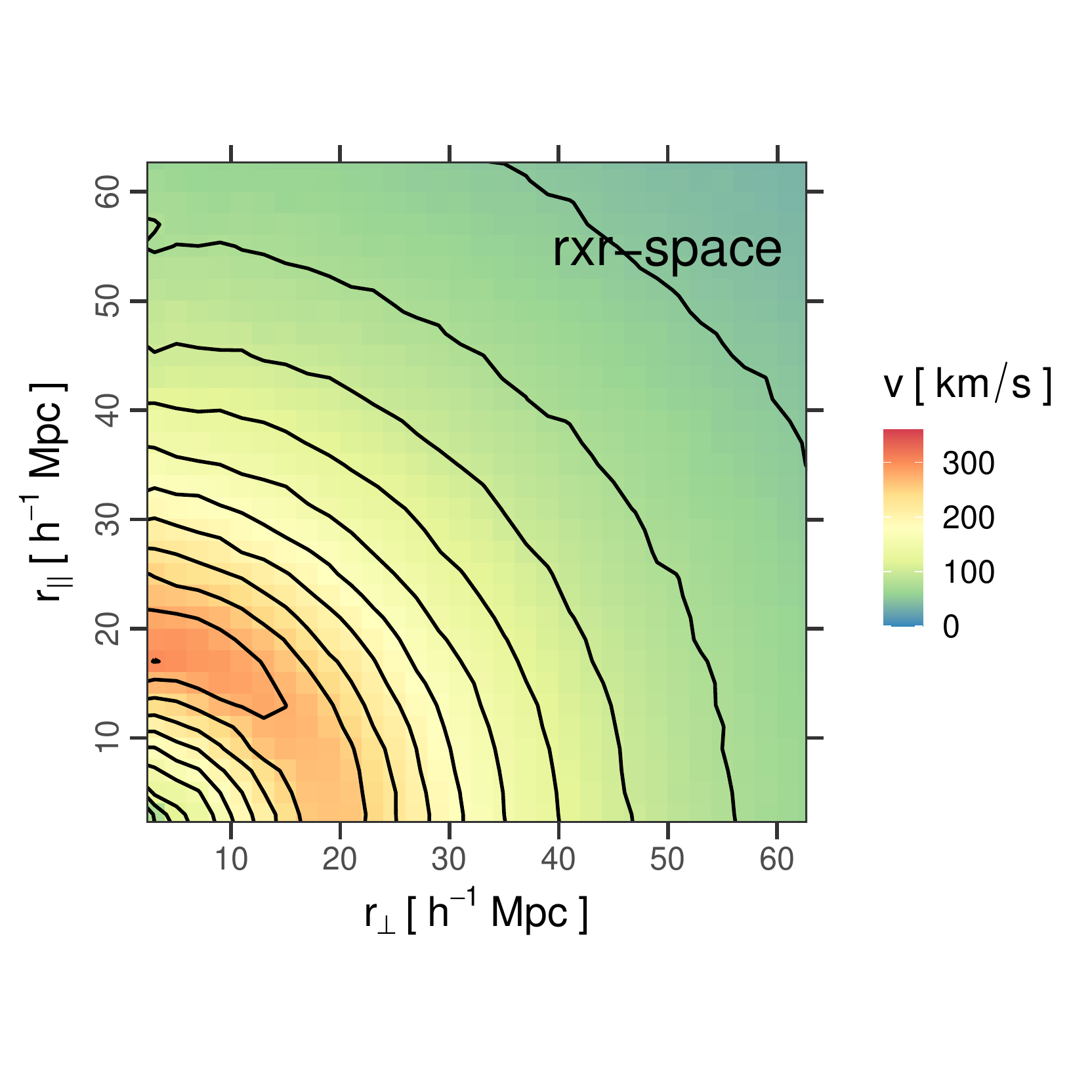}
    \caption{
    Density contrast (left-hand panel) and velocity (right-hand panel) maps in $\rxr$-space corresponding to the overall $r$-space sample.
    The opposite behaviour observed in Figs.~\ref{fig:map2d_dens} and \ref{fig:map2d_vel} does not cancel because OE voids are many more than UE voids.
    The tails of the sample distribution in Fig.~\ref{fig:samples} are the responsible for an additional distortion pattern observed on correlations not previously taken into account, which is due to the intrinsic ellipticity of voids (ellipticity-RSD).
    } 
    \label{fig:map2d}
\end{figure*}

Fig.~\ref{fig:map2d} shows the $\rxr$-space density (left-hand panel) and velocity (right-hand panel) fields corresponding to the overall $r$-space sample.
Note that the anisotropies are still present, the opposite behaviour of the tails does not cancel.
Particularly, the behaviour of the left tail prevails over the right tail.
This is because the former contains many more voids than the latter.
Therefore, although the spherical averaged statistics erase the ellipsoidal nature of voids when estimating the density and velocity profiles, as Fig.~\ref{fig:profiles} shows, it manifests when we calculate correlations. 
This is the reason why the model cannot reproduce correctly the correlation measurements.
We arrive at an important conclusion here: besides the tracer-RSD and void-RSD effects, the intrinsic ellipticity of voids is another source of distortions in the correlation function.
We refer to this effect, particularly to the type of distortions that it generates, with the acronym ellipticity-RSD (or simply e-RSD).
Interestingly, analogous effects have been found in the case of dark matter halos (see for instance \citealt{ersd_clustering_obuljen}).

\subsection{Towards an improved model}
\label{subsec:impact_gsm}

In order to match appropriately the observations, current models for the void-galaxy cross-correlation function must be improved by taking into account the $z$-space effects studied in this work.
By far, the most important aspect is the tracer-RSD+AP correction, i.e. the use of Eq.~(\ref{eq:q_ap_rsd}) to relate to the correct $\rxr$-space statistical properties of the sample, namely, the density and velocity fields.

The remaining deviations are smaller and are attributed to two sources of distortions: void-RSD and ellipticity-RSD.
For the former, the peculiar velocity of voids must be statistically taken into account in the models, as we mentioned in Section~\ref{subsec:impact_vrsd}.
For the latter, a possible solution is to rewrite the GS model by taking into account the elliptical symmetry that the $\rxr$-space density and velocity fields impose, and the connection between them with a generalisation of Eq.~(\ref{eq:velocity}).
Incidentally, there are previous works about the ellipticity of voids and its cosmological importance in the literature \citep{ellipticity_park_lee, ellipticity_bos}.
We leave these topics for a future investigation.

\begin{figure*}
    \includegraphics[width=\columnwidth]{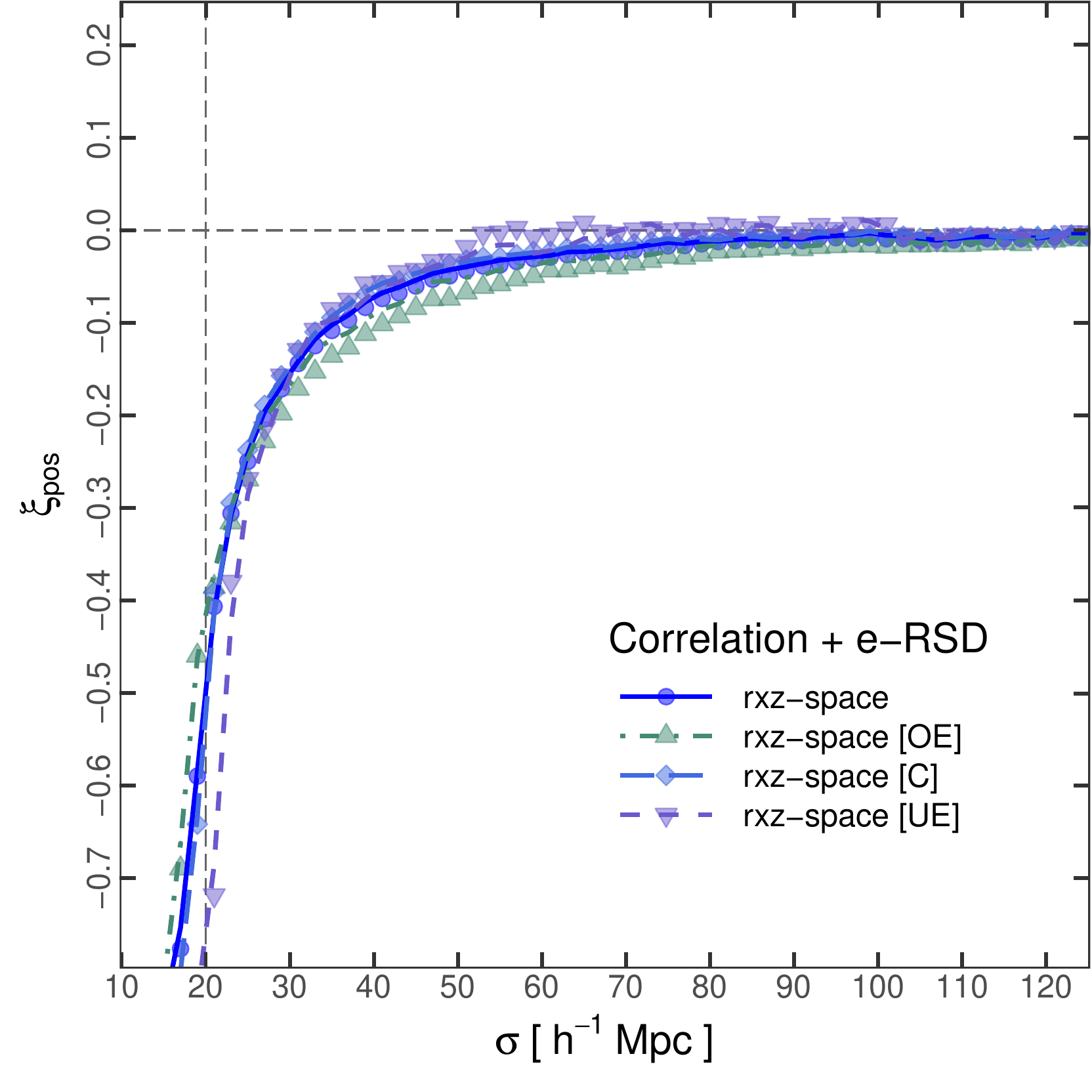}
    \includegraphics[width=\columnwidth]{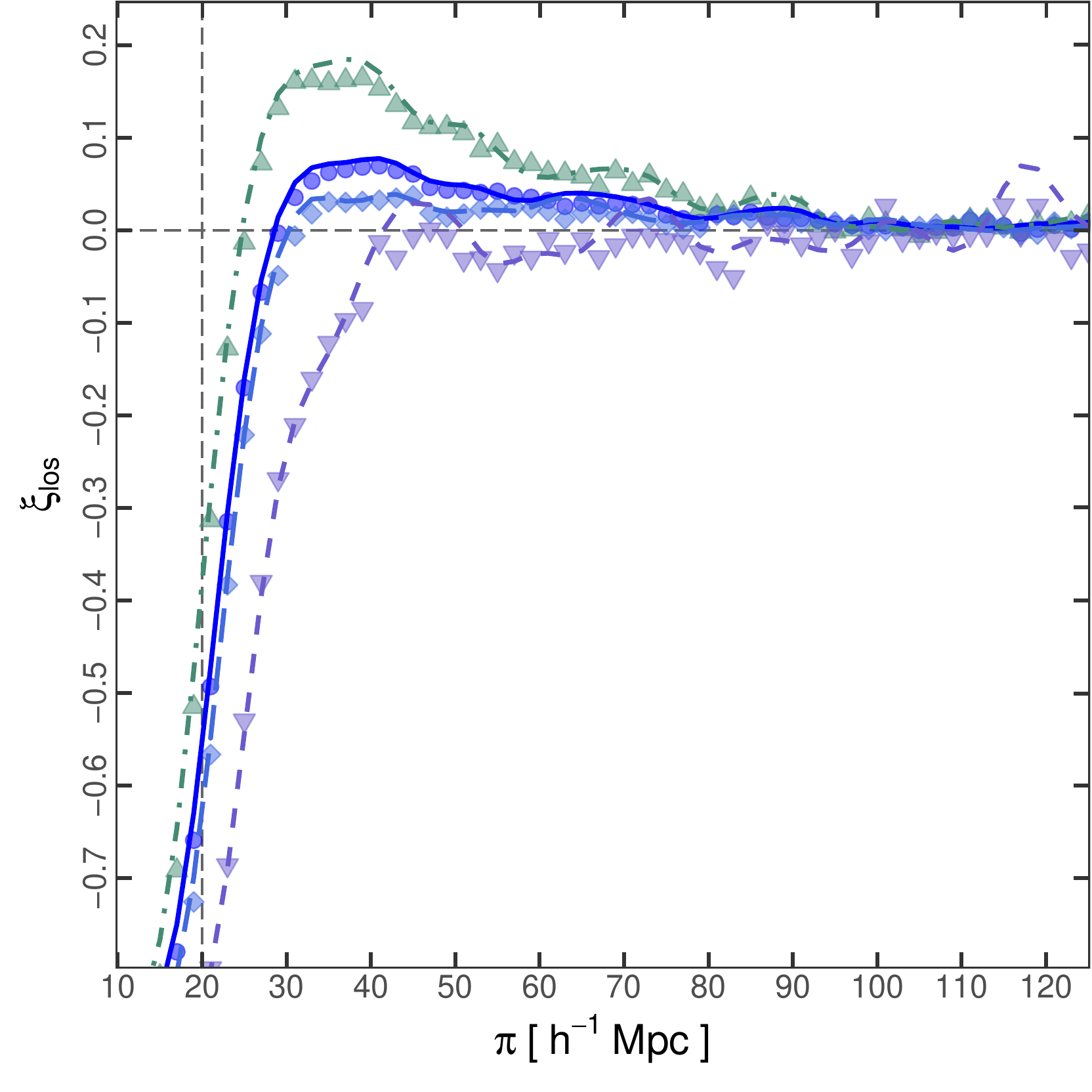}
    \caption{
    Testing the validity of the model by incorporating the elliptical information of the density and velocity maps in $\rxr$-space from Figs.~\ref{fig:map2d_dens} to \ref{fig:map2d} as inputs.
    The $\rxz$-space projected correlations were measured again using a thinner projection range of $10~\hmpc$.
    The measurements are represented with data points, whereas the model predictions with lines.
    Specifically, the samples involved are the following: overall $r$-space sample (blue dots + solid lines), subsample OE (green up-triangles + dot-dashed lines), subsample C (light-blue diamonds + long-dashed lines) and subsample UE (purple down-triangles + dashed lines).
    }
    \label{fig:egsm}
\end{figure*}

In the meantime, and as a first approach, we tested our model again by incorporating the two-dimensional information of the $\rxr$-space density and velocity fields from Fig.~\ref{fig:map2d}.
We measured again the $\rxz$-space projected correlations, but using a thinner projection range: $\mathrm{PR}=10~\hmpc$, which allows us to capture more effectively the behaviour along both directions. 
This is shown in Fig.~\ref{fig:egsm} with blue dots.
Incidentally, the binning step used here is $\delta \sigma = \delta \pi = 2~\hmpc$.
Instead of using a single $\rxr$-space density profile as input in the model, we used two profiles: one suitable for the POS correlation, $\xiposs$, and one suitable for the LOS correlation, $\xiloss$.
Both of them were obtained projecting $\xi(r_\perp,r_\parallel)$ towards the POS and LOS using the same $\rm PR$, respectively.
In the figure, the model prediction is represented with a blue solid line.
Note that it matches the observed data points very well at all scales.
 
For completeness, we also performed this analysis applied to the subsamples defined at the beginning: OE (green up-triangles + dot-dashed lines), C (light-blue diamonds + long-dashed lines) and UE (purple down-triangles + dashed lines).
This is also shown in Fig.~\ref{fig:egsm}.
As before, the model recovers the data points remarkably well at all scales.
These results demonstrate that the GS model is still robust in this case.

\section{Conclusions}
\label{sec:conclusions}

Our standard picture of dynamical and geometrical distortions around voids is incomplete.
Traditionally, we have focused only on the spatial distribution of galaxies.
The truth is that the RSD and AP effects also have an impact on the void identification process itself, affecting intrinsic void properties, such as their number, size and spatial distribution.
This problematic is important when designing cosmological tests, since these void systematics generate additional deviations on the observations leading to biased cosmological constraints if they are not taken into account properly.
This is particularly important in view of the extraordinary precision achievable with modern spectroscopic surveys, such as BOSS, eBOSS and the future DESI, Euclid and HETDEX, which will cover a volume and redshift range without precedents.

One approach is to use a reconstruction technique.
While this method has proved to be robust in recovering the real-space statistical properties of voids, such as their density and velocity profiles, and achieving unbiased cosmological constraints, it also has some disadvantages.
For instance, it is computationally expensive, quite redundant and the valuable cosmological and dynamical information contained in the $z$-space void systematics is not fully exploited.
In Paper~I, we proposed an alternative approach, namely, to find a physical connection between the identification of voids in real and redshift space using a spherical void finder.
A fundamental aspect is that voids above the shot-noise level are almost conserved under the $z$-space mapping, therefore, it is valid to assume void number conservation.
In this context, the differences between the statistical properties between both void populations can be explained by means of three independent effects: the tracer-RSD expansion effect, the AP volume effect, and the void-RSD off-centring effect.
We provided a theoretical framework to describe physically these effects from dynamical and cosmological considerations, and the impact they have on the void size function as a cosmological test.

This work is the continuation of the analyses presented in Paper~I.
Here, we focus on the void-galaxy cross-correlation function.
We adopted the methodology of C19 by analysing two perpendicular projections of the correlation function with respect to the line-of-sight direction.
Therefore, the main conclusions of this paper supplement those of C19 and Paper~I, and can be summarised in the following statements.

\begin{enumerate}

\item[1.]
\textit{Impurity of the sample.}
The impurity of a sample regarding non-bijective voids has a negligible impact on correlation measurements.
This reinforces the fact that void number conservation is a valid assumption.

\item[2.]
\textit{Configurations.}
We measured correlations in different configurations of the spatial distribution of haloes and voids.
Measurements made with $z$-space voids and $z$-space haloes constitute the $\zxz$-space configuration, which represent possible observational measurements.
Similarly, measurements made with $r$-space voids and $z$-space haloes constitute the hybrid $\rxz$-space configuration, where current RSD models are defined to work.
Finally, measurements made with $r$-space voids and $r$-space haloes constitute the $\rxr$-space configuration, free of RSD and AP distortions.

\item[3.]
\textit{Expansion and AP volume effects.}
It is fundamental to provide models with the correct $\rxr$-space statistics of a void sample, namely, the density and velocity fields.
This can be largely achieved by means of a tracer-RSD+AP correction of void radii with Eq.~(\ref{eq:q_ap_rsd}).
The remaining deviations between observations and the model prediction are smaller and caused by the following two sources.

\item[4.]
\textit{Off-centring effect.}
This effect is responsible for an additional distortion pattern due to void dynamics (void-RSD), which can be largely reduced by correcting the centre positions with Eq.~(\ref{eq:vrsd}).
This is the first time that this type of distortions is detected and quantified.

\item[5.]
\textit{Void ellipticity.}
Voids are typically ellipsoidal, but oriented randomly in space.
Therefore, this ellipticity has no significant impact on the statistical properties of a stacking of a complete sample of voids.
However, when a sample is selected in $z$-space, the $r$-space radius of their counterparts distribute in a complex way, covering an extended range of scales.
The tracer-RSD correction predicts the region of completeness, where the ellipticity is not important.
Nevertheless, the tails of the distribution have an appreciable effect on correlations.
They are composed of special voids: over-expanding voids elongated along the POS direction, and under-expanding voids elongated along the LOS direction, responsible for an additional distortion pattern not previously taken into account (ellipticity-RSD).

\item[6.]
\textit{Towards an improved model.}
The tracer-RSD+AP correction is the most important modification needed in models.
Although the remaining deviations are smaller, they have a significant impact.
The void-RSD effect can be corrected for by incorporating information about the void velocity distribution.
This is the connection needed between the $\rxz$- and $\zxz$-space configurations.
Regarding the ellipticity-RSD effect, a possible solution is to rewrite the Gaussian streaming model by taking into account the elliptical symmetry that the $\rxr$-space density and velocity fields impose, and deepening our understanding about the connection between them.
We leave these topics for a future investigation.
With a simplified test, we showed that the GS model can still be robust even in this case.

\item[7.]
\textit{Comparing the void size function and the void-galaxy cross-correlation function.}
From C19, Paper~I and this work, we can conclude that the void size function is affected by two types of distortions: tracer-RSD and AP, whereas the correlation function, on the other hand, by five types of distortions: tracer-RSD, AP, mixture of scales, void-RSD and ellipticity-RSD.

\item[8.]
\textit{Potential of the spherical void finder.}
The simplicity of the spherical void finder allows us to explain naturally the redshift-space effects in voids.
This is because the method finds the largest non-overlapping sphere in each void region, leading to a well defined centre and radius that describe the scale size at a given underdensity threshold.

\end{enumerate}

C19, Paper~I and the present paper are intended to be complementary and part of a global and consistent analysis.
For this reason, we have used the same data set obtained from the Millennium XXL simulation.
However, this data set represents a high-density population of tracers that allows for the selection of relatively small voids ($20-25~\hmpc$).
Current and forthcoming surveys around $z=1$ will have significantly sparser tracer densities.
\citet{rsd_hawken2} show that sparse surveys can be dominated by spurious voids, which could place difficulties when it comes to measure the correlation function and constrain the cosmological parameters.
A more realistic sample would be required to make feasibility studies for these surveys.
With this in mind, we have recently started a similar study based on BOSS data (Correa et al. in preparation), finding promising results.
The POS and LOS correlation functions can be measured with high accuracy as required by current cosmological experiments.
In particular, there is a high signal-to-noise ratio to detect and examine all the effects explained in this work.

As a final reflection, we want to highlight that, in addition to the cosmological importance of considering the void systematics to obtain unbiased constrains, they are also important for large-scale structure studies per se, since they encode valuable information about the nature of cosmic voids regarding their structure and dynamics, and more generally, of the Universe at the largest scales.
Furthermore, some of these effects can constitute cosmological probes by themselves.
For instance, \citet{ellipticity_park_lee} show that the ellipticity distribution of voids constitutes a cosmological test.



\section*{Acknowledgements}

This work was partially supported by the Consejo de Investigaciones Cient\'ificas y T\'ecnicas de la Rep\'ublica Argentina (CONICET) and the Secretar\'ia de Ciencia y T\'ecnica de la Universidad Nacional de C\'ordoba (SeCyT).
This project has received financial support from the European Union's Horizon 2020 Research and Innovation programme under the Marie Sklodowska-Curie grant agreement number 734374 - project acronym: LACEGAL.
This research was also partially supported by the Munich Institute for Astro- and Particle Physics (MIAPP) which is funded by the Deutsche Forschungsgemeinschaft (DFG, German Research Foundation) under Germany´s Excellence Strategy – EXC-2094 – 390783311.
CMC acknowledges the hospitality of the Max Planck Institute for Extraterrestrial Physics (MPE), where part of this work has been done.
ANR acknowledges the financial support of the Agencia Nacional de Promoci\'on Cient\'ifica y Tecnol\'ogica (ANPCyT, PICT 2016-1975).
NP was supported by FONDECYT Regular 1191813, and ANID project Basal AFB-170002, CATA.
REA acknowledges the support of the ERC-StG number 716151 (BACCO).
Numerical calculations were performed at the computer clusters from the Centro de C\'omputo de Alto Desempe\~no de la Universidad Nacional de C\'ordoba (CCAD, http://ccad.unc.edu.ar).
Plots were made with the ggplot2 package \citep{ggplot2} of the R software \citep{R} and post-processed with Inkscape (https://inkscape.org).
CMC would like to specially thank Daniela Taborda for her unconditional support.


\section*{Data Availability}

The data underlying this article will be shared on reasonable request to the corresponding author.


\bibliographystyle{mnras}

\input{voids_zspace.bbl}


\bsp	
\label{lastpage}
\end{document}